\documentclass{aa}
\usepackage{array}
\usepackage{graphicx}
\usepackage{booktabs}
\usepackage[varg]{txfonts}
\usepackage{xcolor}
\usepackage[draft]{hyperref}
\usepackage{natbib}
\bibpunct{(}{)}{;}{a}{}{,}
\defcitealias{ilin2019}{Paper~I}

\begin{document}
%LINEBREAKS FOR APPENDIX
\newcolumntype{L}[1]{>{\raggedright\let\newline\\\arraybackslash\hspace{0pt}}m{#1}}
%LINEBREAK FOR APPENDIX END

   \title{Flares in Open Clusters with K2.}

   \subtitle{II. Pleiades, Hyades, Praesepe, Ruprecht 147, and M67}

   \author{Ekaterina Ilin
          \inst{1,2}, Sarah J. Schmidt\inst{1},
          Katja Poppenh\"ager\inst{1,2},
          James R. A. Davenport\inst{3},
           Martti H. Kristiansen\inst{4,5},
           Mark Omohundro\inst{6}
          }

   \institute{Leibniz-Institute for Astrophysics Potsdam (AIP), An der Sternwarte 16, 14482 Potsdam, Germany\\\email{eilin@aip.de},\and
   Institute for Physics and Astronomy, University of Potsdam, Karl-Liebknecht-Str. 24/25, 14476 Potsdam, Germany\and
              Department of Astronomy, University of Washington, Seattle, WA 98195, USA\and
              Brorfelde Observatory, Observator Gyldenkernes Vej 7, DK-4340 T\o ll\o se, Denmark\and
              DTU Space, National Space Institute, Technical University of Denmark, Elektrovej 327, DK-2800 Lyngby, Denmark \and
              Citizen Scientist, c/o Zooniverse, Department of Physics, University of Oxford, Denys Wilkinson Building, Keble Road, Oxford, OX1 3RH, UK
             }

   \date{Received August 17, 2020; accepted October 9, 2020}

% \abstract{}{}{}{}{}
% 5 {} token are mandatory

  % context heading (optional)
  % {} leave it empty if necessary
\abstract{Magnetic fields are a key component in the main sequence evolution of low mass stars. Flares, energetic eruptions on the surfaces of stars, are an unmistakable manifestation of magnetically driven emission. The occurrence rates and energy distributions of flares trace stellar characteristics such as mass and age. But before flares can be used to constrain stellar properties, the flaring-age-mass relation requires proper calibration.}{This work sets out to quantify flaring activity of independently age-dated main sequence stars for a broad range of spectral types using optical light curves obtained by the Kepler satellite.}{Drawing from the complete K2 archive, we searched 3435 $\sim 80$\,day long light curves of 2111 open cluster members for flares using the open-source software packages K2SC to remove instrumental and astrophysical variability from K2 light curves, and AltaiPony to search and characterize the flare candidates.}{We confirmed a total of 3844 flares on high probability open cluster members with ages from zero age main sequence (Pleiades) to 3.6 Gyr (M67). We extended the mass range probed in the first study of this series to span from Sun-like stars to mid-M dwarfs. We added the Hyades (690 Myr) to the sample as a comparison cluster to Praesepe (750 Myr), the 2.6 Gyr old Ruprecht 147, and several hundred light curves from the late K2 Campaigns in the remaining clusters. We found that the flare energy distribution was similar in the entire parameter space, following a power law relation with exponent $\alpha\approx 1.84-2.39$. }{We confirmed that flaring rates declined with age, and declined faster for higher mass stars. Our results are in good agreement with most previous statistical flare studies. We found evidence that a rapid decline in flaring activity occurred in M1-M2 dwarfs around Hyades/Praesepe age, when these stars spun down to rotation periods of about 10\,d, while higher mass stars had already transitioned to lower flaring rates, and lower mass stars still resided in the saturated activity regime. We conclude that some discrepancies between our results and flare studies that used rotation periods for their age estimates could be explained by sample selection bias toward more active stars, but others may hint at limitations of using rotation as an age indicator without additional constraints from stellar activity.}
             \titlerunning{Flares in Open Clusters with K2. II.}
\authorrunning{E. Ilin et al.} 
   \maketitle

\section{Introduction}
Flares are explosions that occur on the surfaces of practically all low mass stars down to the very bottom of the main sequence. We know that flares are magnetic re-connection events that lead to a change in field line topology and subsequent energy release~\citep{priest_magnetic_2002}. We can observe flares in nearly all electromagnetic bands, from radio
to X-ray, and on all stars that possess a convection zone, from late F type stars to ultracool dwarfs~\citep{schaefer2000,benz2010,gizis2013}. 
\\
Stellar flares on cool stars can enhance the optical flux by multiple orders of magnitude within minutes or seconds~\citep{haisch1991, schmidt2019}, and release energies up to \mbox{$\sim10^{37}$ erg}~\citep{maehara2012, davenport_kepler_2016}. They typically exhibit blackbody emission at about 10\,000 K~\citep{hawley1992, kowalski2013}, which is significantly hotter than the photospheres of these stars (2500 $-$ 6500 K). The high intensity, strong contrast, and broad energy distributions allow us to measure magnetic activity as traced by flares for a great variety of time resolved observations.
\\
Statistical flare studies were pioneered from the ground~\citep{lacy_uv_1976}, but it was not until space missions like Kepler~\citep{koch2010} that investigating stellar ensembles that were not pre-selected for their activity was possible~\citep{walkowicz2011}. Today, the ground based all-sky surveys ASAS-SN~\citep{shappee2014} and Evryscope~\citep{law2015}, and also the Transiting Exoplanet Survey Satellite~(TESS,~\citealt{ricker2014}) are following in the footsteps of Kepler~(see works by~\citealt{schmidt2019,rodriguez2020,howard2019, howard2020, guenther2020, feinstein2020}). Statistical studies of stellar flaring activity can help us understand the underlying physical processes of flares~\citep{benz2010}, the nature and strength of stellar magnetic fields~\citep{berger2006, odert2017}, starspots~\citep{davenport_flaresandspots_2015, notsu2019, howard2020}, how flares relate to stellar angular momentum evolution~\citep{mondrik2019, howard2020}, and how they affect the atmospheres of exoplanets~\citep{lecavelier_flareescape_2012, loyd_mflaresplanetsfuv_2018, tilley_repeated_flare_2019, howard2019}.
\\
The present study has a focus on the relation between flares and the age of low mass stars on the main sequence. To this end, we set out to quantify how flaring activity changes as a function stellar mass and age. 
\\
In cool main sequence stars, the fraction of stars found flaring in the optical is known to increase down to spectral type M5~\citep{yang_flaring_2017, chang2020}. Even beyond M5, and down to L dwarfs, flaring is ubiquitous~\citep{stelzer2006, robrade2010, gizis2013, schmidt2015, schmidt2016, paudel2018, paudel2020}. In fact, the prototype flare star, UV Ceti, is an M6 dwarf~\citep{kirkpatrick1991}.  
Flaring activity also decays with age, but stellar ages can only be determined indirectly. Using galactic latitute as a proxy for age, flaring activity appeared to decline with higher galactic latitude, that is, older age, for M dwarfs~\citep{hilton2010, walkowicz2011, howard2019}. In gyrochronology, fast rotation indicates young age~\citep{barnes_rotational_2003}, and slows down as the star ages. So the relation between rotation and flaring can be used to calibrate a flaring-age relation~\citep{davenport2019}. 
\\
In M dwarfs, the fraction of stellar luminosity emitted in flares is correlated with the fraction emitted in H$\alpha$ in quiescence~\citep{yang_flaring_2017}. Quiescent H$\alpha$ emission was also explored as an age tracer in the past~\citep{soderblom_chromospheric_1991, pace_chromospheric_2013, lorenzo-oliveira_age-mass-metallicity-activity_2016}. More recent work focused on stars older than a gigayear, and calibrated the H$\alpha$-age relation using isochronal~\citep{lorenzooliveira2018}, and asteroseismic ages~\citep{booth2017, booth2020}, suggesting that an age calibration of flaring activity is possible.
\\
While absolute stellar ages are difficult to assess, they can be differentially controlled for in coeval groups of stars. Flaring-age studies in binaries showed consistent activity for both components in the majority of targets~\citep{lurie2015, clarke_flare_2018}. Open clusters are also coeval groups of stars with well-determined isochronal ages that have been used as a laboratory for flare studies on stars with a fixed age~\citep{mirzoyan1993, chang2015}. \citet{ilin2019}~(hereafter \citetalias{ilin2019}) investigated the flaring activity of late-K to mid-M dwarfs in three open clusters, the Pleiades, Praesepe, and M67, using K2 time domain photometry. We analyzed flare frequency distributions (FFDs) broken down by the stars' effective temperatures $T_\mathrm{eff}$ and ages. We found that flaring activity declined both with increasing mass and age, and that the trend was more pronounced for higher mass stars. 
\\
This study extends the results in \citetalias{ilin2019} to the age of Ruprecht 147 (2.6 Gyr), and both higher and lower masses than in the previous study. We used the now complete K2 data set, supplemented all three open clusters in~\citetalias{ilin2019} with improved versions of already treated light curves in Campaigns 4 and 5, and added light curves from later Campaigns. The light curve catalog, and the determination of cluster membership and effective temperature for the investigated stars are detailed in Section \ref{sec:data}. We describe how we used a semi-automated flare finding procedure, how we estimated flare energies, and how we parametrized the statistical properties of flares in Section \ref{sec:methods}. We present our findings in Section \ref{sec:results}. We place our results in the context of recent flare studies, and reflect on the power law nature of FFDs in Section \ref{sec:discussion}. Recently, \citet{davenport2019} proposed an empirical parametrization of the flaring-mass-age relation based on FFDs of stars with gyrochronologically determined ages, which we also put to test in the discussion. The summary and conclusions can be found in Section \ref{sec:summary}.
\section{Observations and data analysis}
\label{sec:data}
\begin{figure}[t]
            \includegraphics[width=\hsize]{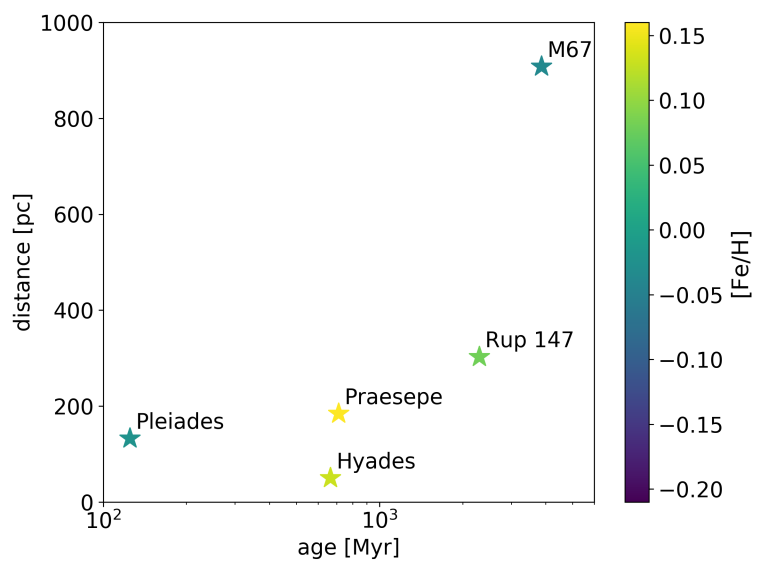}
         \caption{Distance as a function of age for open clusters (star symbols) that we searched for flares, colored by [Fe/H]. Age, distance, and [Fe/H] are approximate values from a compilation of existing literature, see Appendix \ref{app:cluster_params}.}
         \label{fig:OCs}
   \end{figure}

This work is based on K2 long cadence light curves with an integration time of 30 minutes that were provided by the Kepler archives hosted at the Barbara A. Mikulski Archive for Space Telescopes (MAST)~(Section~\ref{sec:sec:k2lc}). To select a sample with independently derived ages we began by gathering open cluster membership information from the literature~(Section~\ref{sec:sec:ocmem}). An overview over the cluster sample is presented in Table \ref{tab:data_clusters} and illustrated in Figure \ref{fig:OCs}. To characterize the stars in the resulting catalog for the analysis that followed, we used multiband photometry from several all-sky surveys to assign $T_\mathrm{eff}$~(Fig. \ref{fig:teff_spread}) and determine stellar radii~(Fig. \ref{fig:teff_radius}) using empirical relations, and eventually calculated stellar luminosities in the Kepler band~(Section~\ref{TeffRL}).

%-------------------------------------------
%--OPEN CLUSTERS DISTANCE AND AGE
%-------------------------------------------

\subsection{K2 light curves}
\label{sec:sec:k2lc}
The Kepler~\citep{koch2010} spacecraft finished its follow-up mission K2~\citep{howell_k2_2014} in September 2018, after having completed nearly 20 80-day observing campaigns. With the full archive at hand, we selected and analyzed a total of 3435light curves of high-confidence members of five open clusters. Each light curve contained up to $80$ uninterrupted days of $30$ min cadence observations in white light~($4,200-9,000\,\mathring{A}$).
\\
As K2 was conducted on the two-wheeled Kepler satellite, it was subjected to substantial drift motion (spacecraft roll,~\citealt{van_cleve_thats_2016}) and had an overall reduced pointing accuracy. To mitigate these effects, various solutions were developed~\citep{vanderburg_k2sff_2014, aigrain_k2sc_2016, luger_everest_2016, luger_everst_2018}. We used the K2 Systematics Correction (\texttt{K2SC}) pipeline~\citep{aigrain_k2sc_2016} with minor modificiations to de-trend the 36th data release of the K2 data products. This data release was final, and included a uniform, global reprocessing of most K2 campaigns using an improved data reduction pipeline\footnote{\url{https://keplerscience.arc.nasa.gov/k2-uniform-global-reprocessing-underway.html}}.
\subsection{Open cluster membership}
\label{sec:sec:ocmem}
For each open cluster, we obtained membership information from a selection of catalogs which we cross-matched on right ascension and declination within 3 arcsec against the full K2 archive. One part of the membership catalogs provided membership probabilities~\citep{douglas_praesepe_hyades_2014, cantat_gaudin_2018, olivares_pleiades_2018, reino_hyades_2018, gao_m67mem_2018, olivares_ngc6774_2019}. For the other part no probability was quoted~\citep{rebull_pleiadesrot_2016, douglas_poking_2017, gaia_dr2_2018_hrd}, or qualitative classifiers were given~\citep{curtis2013, gonzalez_m67mem_2016,rebull_praesepe_2017}. In the latter cases we assigned approximate probabilities anchored to the set threshold for inclusion into our final sample~(Appendix \ref{app:memberships}). Absence in a catalog did not decrease the likelihood of membership, as each catalog had different selection biases which we did not address in this study. We set the threshold mean membership probability $p$ for a target in our sample to 0.8. 
\\
All in all, we identified 2111low mass stars in five open clusters spanning ages from zero age main sequence (ZAMS) to roughly solar. Table \ref{tab:data_clusters} provides an overview over the final sample of stars and flares, and the adopted open cluster properties. Membership probability histograms of the final sample are displayed in Figure \ref{fig:app:memberships} in Appendix \ref{app:memberships}. A literature overview of age, distance, and metallicity determinations for the open clusters from which we adopt the values for this study is given in Table \ref{tab:app:oc_parameters} in Appendix \ref{app:cluster_params}\footnote{The analysis code for this section can be found at \url{https://github.com/ekaterinailin/flares-in-clusters-with-k2-ii} in the "Membership\_Matching" directory}. Below, we provide additional details on the final samples of each cluster.
\begin{table*}
\caption{Open clusters.}
\label{tab:data_clusters}
\centering
\begin{tabular}{lccccccccr}
\hline\hline
          &  d [pc] &  stars &   LCs &  flares &    campaigns &                        age [Myr] &         [Fe/H] \\
\hline
 Pleiades &   135.6 [1] &    741 &   741 &    1583 &            4 &     $135\left(_{25}^{25}\right)$ [2]&  $-0.04(0.03)$ [3] \\
   Hyades &    46.0 &    170 &   179 &     402 &      4, 13 &   $690\left(_{100}^{160}\right)$ [4] &   $0.13(0.02)$ [5]\\
 Praesepe &   185.5 [1]&    913 &  1965 &    1849 &  5, 16, 18 &       $750\left(_{7}^{3}\right)$ [6]&   $0.16(0.00)$ [5]\\
  Ruprecht 147 &   305.0 [1]&     53 &    53 &       9 &            7 &  $2650\left(_{380}^{380}\right)$ [7]&   $0.08(0.07)$ [8] \\
      M67 &   908.0 [9]&    234 &   497 &       1 &  5, 16, 18 &    $3639\left(_{17}^{17}\right)$ [6]&  $-0.10(0.08)$ [3]\\
\hline

\end{tabular}

\tablefoot{The values for age, [Fe/H], and distance $d$ are approximate values arrived at by a comparison of existing literature (see Table~\ref{tab:app:oc_parameters} in the Appendix). The uncertainties are noted in parentheses. "stars" denotes the number of cluster members with membership probability $p>0.8$. "LCs" and "campaigns" are the numbers of available light curves, and the K2 campaigns during which they were observed, respectively. "flares" is the number of confirmed flares found in each cluster. The quoted Hyades distance is the mean Gaia DR parallax distance. \tablebib{
 [1] \citet{cantat_gaudin_2018}, [2] \citet{gossage2018}, [3] \citet{conrad2014}, [4] \citet{gaia_dr2_2018_hrd}, [5] \citet{netopil_metallicities_2016}, [6] \citet{bossini2019}, [7] \citet{torres2018}, [8] \citet{bragaglia2018}, [9] \citet{dias_fitting_2012}}}.
\end{table*}
%---------------------------------------------------------------
\subsubsection{Pleiades}
The Pleiades, a nearby ZAMS cluster with age estimates ranging from 87~Myr to 141~Myr~\citep{bell_pre-main-sequence_2012, scholz2015, dahm_reexamining_2015, yen2018, gossage2018, bossini2019}, was observed in Campaign 4, and treated in \citetalias{ilin2019}. Here, we used 741 reprocessed K2 light curves from this open cluster, and our methodological improvements to \citetalias{ilin2019} to repeat the analysis. We revisited the members from~\citet{rebull_pleiadesrot_2016}, which were used in \citetalias{ilin2019}, and merged the catalog with lists of members determined by~\citet{olivares_pleiades_2018, gaia_dr2_2018_hrd}; and~\citet{cantat_gaudin_2018}.
\subsubsection{Hyades}
The Hyades are a $625-690$~Myr old open cluster~\citep{perryman1998,salaris_age_2004, gossage2018, gaia_dr2_2018_hrd} that was observed during Campaigns 4 and 13 with K2.  We merged membership tables obtained from \citet{douglas_praesepe_hyades_2014, reino_hyades_2018}; and \citet{gaia_dr2_2018_hrd} to select 170 high confidence members, of which 9 were observed in both K2 campaigns. 
%---------------------------------------------------------------
\subsubsection{Praesepe}
Compared across different studies, Praesepe ($590-830$~Myr,~\citealt{salaris_age_2004, boudreault_astrometric_2012, scholz2015, yen2018, gossage2018, bossini2019}) is a little older than the Hyades on average. The open cluster was observed during Campaign 5, and was also treated in~\citetalias{ilin2019}. It was observed again during campaigns 16 and 18. We revisited the membership catalog from~\citet{douglas_praesepe_hyades_2014}, and matched it to the members identified in \citet{douglas_poking_2017, rebull_praesepe_2017,cantat_gaudin_2018}; and ~\citet{gaia_dr2_2018_hrd}. This was the richest sample with over 900 stars and nearly 2000 light curves.
%---------------------------------------------------------------
\subsubsection{Ruprecht 147}
Ruprecht 147 is an open cluster with estimated ages from 2.0~Gyr to 3.1~Gyr~\citep{curtis2013,scholz2015,gaia_dr2_2018_hrd,torres2018} observed during Campaign 7 with K2. We used the mean membership probabilities obtained from ~\citet{curtis2013, cantat_gaudin_2018, olivares_ngc6774_2019}, and~\citet{gaia_dr2_2018_hrd} to identify the 53 most likely members with K2 light curves as described above.
%---------------------------------------------------------------
\subsubsection{M67}
M67 is an old open cluster that has 234 members which were observed during campaigns 5, 16 and 18. Its age is about solar, with estimates ranging from 3.4~Gyr to 4.3~Gyr~\citep{salaris_age_2004, onehag2011, dias_fitting_2012, scholz2015, barnes_rotation_2016, bossini2019}. The relatively large distance of about 900 pc~\citep{dias_fitting_2012} limited the number of observable low mass stars in this otherwise rich cluster. We did not find any flares in M67 in Campaign 5~\citepalias{ilin2019} light curves of members identified by \citet{gonzalez_m67mem_2016}. Campaigns 16 and 18 delivered both additional observations, and new targets to our previous sample, yielding almost 500 light curves for this open cluster. We merged the members from \citet{gonzalez_m67mem_2016} with a recent study based of Gaia DR2 data~\citep{gao_m67mem_2018}.
%---------------------------------------------------------------
\subsection{Effective temperatures, stellar radii, and luminosities}
\label{TeffRL}
To study the flaring-age-mass relation, we needed to determine effective temperatures ($T_\mathrm{eff}$,~Section \ref{sec:sec:sec:teff}), and the luminosities in the Kepler band ($L_\mathrm{Kp,*}$,~Section \ref{sec:lum}) of the investigated stars. $L_\mathrm{Kp,*}$ was required to determine the flare energies, and $T_\mathrm{eff}$ was used as a proxy to stellar mass. Since our catalog of stars spanned a wide variety of observables, no reliable Gaia parallaxes were available for a significant fraction of the targets, no single photometric survey covered all stars with broadband observations, and there was no emprical relation available that could have been used to derive $T_\mathrm{eff}$ for the entire range of spectral types. Empirical color-temperature relations (CTR) suffer from systematic errors that stem both from the different methods applied, and from sample selection biases. We therefore used several empirical relations in their appropriate ranges to obtain $T_\mathrm{eff}$ from each, and draw a more reliable mean. Targets that were lacking sufficient photometric data to derive $T_\mathrm{eff}$, or that were too hot to be expected to have a convective envelope ($T_\mathrm{eff} \geq 7000\,$K), were removed from the sample. We dropped all targets where the uncertainty on the weighted mean $T_\mathrm{eff}$ was greater than $10\,$\%. Only targets that were assigned a $T_\mathrm{eff}$ were searched for flares. To eventually derive $L_\mathrm{Kp,*}$ from $T_\mathrm{eff}$ and model spectra we also required stellar radii as an intermediate step~(Section~\ref{sec:sec:sec:r}).
\subsubsection{Effective temperatures $T_\mathrm{eff}$}
\label{sec:sec:sec:teff}
\begin{figure*}
		\centering
           \includegraphics[angle=90, width=\hsize]{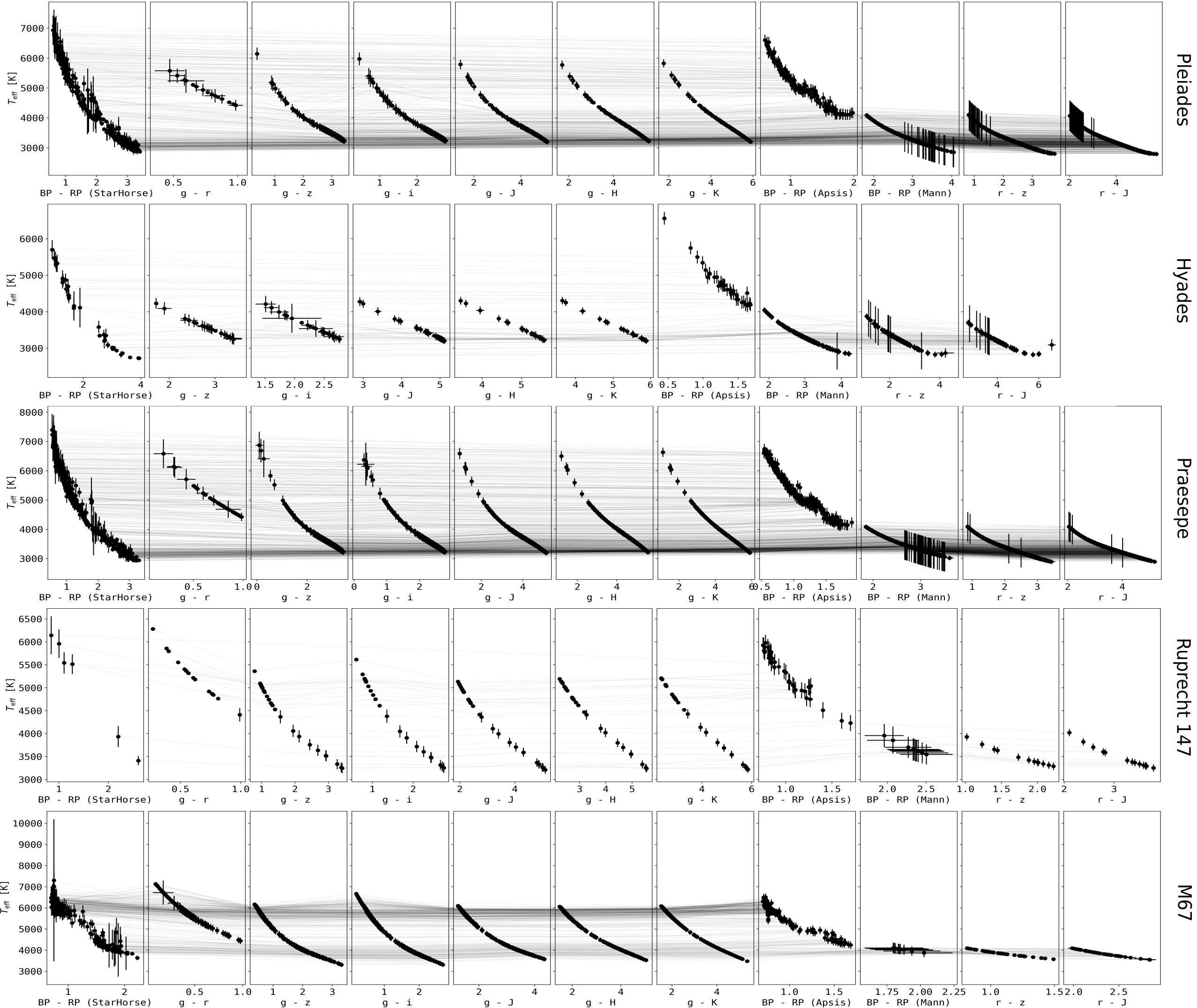}

      \caption{Empricial color-$T_\mathrm{eff}$ relations applied to high probability members of the five open clusters in this study. Individual stars are connected with lines between sub-plots to illustrate differences in the relations. We used PanSTARRS $griz$, 2MASS $JHK$, and Gaia DR2 $BP$ and $RP$ photometry with the following relations: $BP-RP$ (StarHorse, \citealt{anders_starhorse_2019}); $g-r$, $g-i$, $g-z$, $g-J$, $g-H$, $g-K$~\citep{boyajian_stellar_2013}; $BP-RP$ (Apsis, \citealt{andrae_gaiaapsis_2018}); $BP-RP$ (Mann), $r-z$, and $r-J$ \citep{mann_erratum_2016}. The majority of connecting lines show very little slope, indicating that different CTRs are consistent with one another when applied to our sample. As expected, some relations show noticeable systematic differences, especially at the high and low temperature ends of their scopes of validity. We did not count results from Apsis with $T_\mathrm{eff}<4100\,$K because the pipeline systematically overestimated the temperatures for M dwarfs~\citep{andrae_gaiaapsis_2018,kesseli2019}.}
         \label{fig:teff_spread}
   \end{figure*}

We determined $T_\mathrm{eff}$ using broadband photometry from the Two Micron All Sky Survey~(2MASS; \citealt{skrutskie_two_2006}), the Panoramic Survey Telescope and Rapid Response System Data Release 1~(PanSTARRS DR1; \citealt{2016arXiv161205560C}), and Gaia Data Release 2~(Gaia DR2; \citealt{gaia2016, gaia2018}) with the following quality cuts: We required the 2MASS measurements for $J$, $H$, and $K$ to be "A". "A" meant that measurements had $S/N>10$ and $\sigma<0.11$. For PanSTARRS photometry, we required that the \texttt{QF\_OBJ\_GOOD} quality filter flag was set. SDSS and PanSTARRS \textit{ugrizy} bands were similar but not identical. They could be converted using Table 2 in~\citet{finkbeiner_ps1tosdss_2016}. And for Gaia photometry, we cut at \texttt{flux}/\texttt{flux\_error} $\geq 10$ in the $G$, $BP$, and $RP$ bands. Gaia astrometry solutions were only used when the Re-normalized Unit Weight Error (RUWE) was $<1.4$. We also removed foreground stars using Gaia DR2 parallaxes. We corrected the 2MASS and PanSTARRS photometry from the distant clusters M67 and Ruprecht 147 for extinction using the most recent version of the \texttt{dustmaps}~\citep{green_dustmaps_2018} package that provides 3D dust maps derived from 2MASS and PanSTARRS photometry together with Gaia distances~\citep{green2019}. When no Gaia parallax was available we used the cluster median distance instead. If an extinction value was not available for a given star we used the average extinction value of the respective cluster. We accounted for extinction in Gaia $BP$ and $RP$ bands using the reddening $E(B_P-R_P)$ derived from Gaia photometry and parallax from Gaia DR2~\citep{andrae_gaiaapsis_2018}. We dropped targets that were too bright and would saturate the detector (Kepler magnitude $K_\mathrm{p} \leq 9$). Faint targets were not filtered out by a threshold but were removed from the sample when $T_\mathrm{eff}$ did not meet the quality criteria described above.
\\
To determine robust $T_\mathrm{eff}$~(Fig. \ref{fig:teff_spread}), we used CTRs from \citet{boyajian_stellar_2013} and \citet{mann_erratum_2016} (erratum to \citealt{mann_how_2015}), $T_\mathrm{eff}$ derived from Gaia DR2 using the StarHorse algorithm~\citep{queiroz_starhorse_2018}, and $T_\mathrm{eff}$ inferred from Gaia DR2 using the Apsis pipeline~\citep{bailerjones_apsis_2013, andrae_gaiaapsis_2018}.
\\
\citet{boyajian_stellar_2013} determined CTRs from a range of interferometrically characterized stars using $g-r$, $g-i$, $g-z$, $g-J$, $g-H$, and $g-K$ colors from SDSS and Johnson magnitudes for A to K stars. Their sample was centered on solar metallicity, so we constained the use of these CTRs to stars with $-0.25<$\,[Fe/H]$\,<0.25$. Following \citet{boyajian_stellar_2013}, we transformed 2MASS $JHK$ to $J-H$, $H-K$, and $J-K$ in the Johnson system from 2MASS to the Bessell-Brett sytem~\citep{carpenter_color_2001}, and from Bessell-Brett to Johnson using the relations in~\citet{bessell_brett_1988}. 
\\
\citet{mann_how_2015} derived CTRs from absolutely calibrated spectra to which they fitted atmospheric models to obtain $T_\mathrm{eff}$. 
They provided transformations for SDSS/2MASS $r-z$ and $r-J$, or Gaia $BP-RP$ where extra information could be added from metallicity or 2MASS $J-H$. The relations in \citet{mann_how_2015} were only valid if metallicity was sufficiently close to solar, which was satisfied for the open clusters in this paper~(see Table \ref{tab:data_clusters}). 
\\
We supplemented our estimates with $T_\mathrm{eff}$ estimates from \citet{anders_starhorse_2019} who determined distances, extinctions, and various stellar parameters for 137 million stars in Gaia DR2 using the StarHorse pipeline~\citep{queiroz_starhorse_2018}.
\\
Gaia DR2 published $T_\mathrm{eff}$ for over 160 million sources~\citep{gaia2018}. The typical uncertainty was quoted at 324 K, but it was lower for stars above $\sim$4100 K and below $\sim$6700K, so that we adopted 175 K which was above the quoted root-median-squared error in this $T_\mathrm{eff}$ range, and used provided values only in this range, as the pipeline systematically overestimated the temperatures for M dwarfs~\citep{andrae_gaiaapsis_2018,kesseli2019}. 
\\ 
Finally, we plotted color-magnitude diagrams in all aforementioned colors, and flagged outliers from the main sequence. These stars were excluded from the analysis.\footnote{The analysis code for this section can be found at \url{https://github.com/ekaterinailin/flares-in-clusters-with-k2-ii} in the "StellarParameters" directory}.

%---------------------------------------------------------------
% CTRs
%---------------------------------------------------------------

%---------------------------------------------------------------
% Teff - R
%---------------------------------------------------------------

   \begin{figure*}
		\centering
           \includegraphics[width=1.015\hsize]{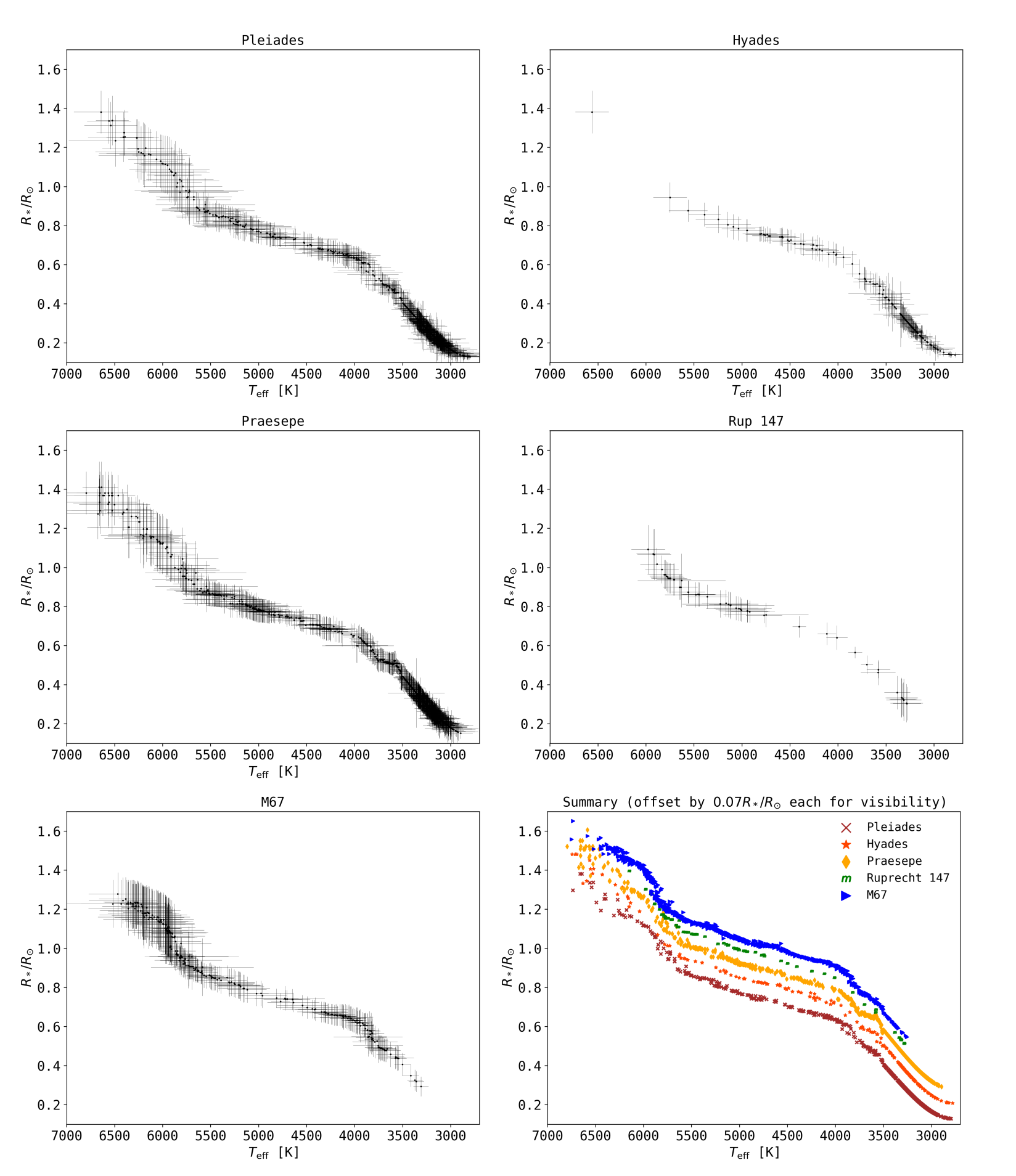}

      \caption{$T_\mathrm{eff}-R_*$ distributions for all clusters in this study. Each panel shows the distribution of high probability members of the open cluster indicated in the panel title. Stellar radii $R_*$ are given in units of solar radii. Uncertainties on $T_\mathrm{eff}$ and $R_*$ (gray error bars) were propagated from the photomety, the color-$T_\mathrm{eff}$ relations, and empirical $T_\mathrm{eff}-R_*$ relations. The uncertainties on $R_*$ were typically $\sim10-15$\,\%, which matches the scatter in the empirical $T_\mathrm{eff}-R_*$ relations. The uncertainties on $T_\mathrm{eff}$ roughly correspond to the size of the temperature bins that we chose in this work~(see, for example, Fig. \ref{fig:powerlawfits_s}). The bottom right panel shows all five cluster samples together for comparison.}
         \label{fig:teff_radius}
         
   \end{figure*}

%---------------------------------------------------------------
   
\subsubsection{Stellar radii}
\label{sec:sec:sec:r}
We used a catalog of empirically characterized stars~\citep{yee_specmatch_2017} and empirical $T_\mathrm{eff}-R_*$ relations derived by~\citet{mann_how_2015,mann_erratum_2016} to obtain $R_*$ from the $T_\mathrm{eff}$ for the stars in our sample (Fig.~\ref{fig:teff_radius}). \citet{yee_specmatch_2017} compiled 404 stars with high-resolution spectra from the literature, and performed their own observations of mid to late K-dwarfs, achieving a coverage of low mass stars from 7000\;K down to 3000\;K. For these stars, the resulting catalog was accurate to 100\;K in
$T_\mathrm{eff}$, 15\;\% in $R_*$, and 0.09\;dex in [Fe/H]. We interpolated between stars from the catalog to our derived $T_\mathrm{eff}$,  and propagated the resulting scatter to the uncertainty in $R_*$ if $T_\mathrm{eff}>3500$ K. For stars with \mbox{$T_\mathrm{eff}< 3500$} K we used $T_\mathrm{eff}-R_*$ relations derived by~\citet{mann_how_2015,mann_erratum_2016}. 
\subsubsection{Stellar spectra and luminosities}
\label{sec:lum}
We assigned spectra to our targets from the SpecMatchEmp\footnote{\url{https://specmatch-emp.readthedocs.io/en/latest/}}~\citep{yee_specmatch_2017} and the FlareSpecPhot\footnote{\url{https://github.com/sjschmidt/FlareSpecPhot}} libraries~\citep{Kirkpatrick2010, Burgasser2007, Burgasser2008, Burgasser2010, Burgasser2004, Cruz2004, Burgasser2006, Rayner2009, Doi2010, Filippazzo2015, Cruz2003, West2011, Bochanski2010,  Bochanski2007, Schmidt2010, schmidt2015, Schmidt2014a, Schmidt2014b, mann_how_2015}. When a spectrum was available for the derived spectral type in FlareSpecPhot, we preferred it over SpecMatchEmp, which was the case for all stars cooler than M0, where we mapped spectral type to effective temperature~\citep{pecaut_intrinsic_2013}. We then combined $R_*$, $T_\mathrm{eff}$, and spectra to projected bolometric luminosities $L_{\mathrm{bol,*}}$, and projected luminosities in the Kepler band $L_{\mathrm{Kp,*}}$ following~\citet{shibayama2013} and~\citetalias{ilin2019}. For $L_{\mathrm{bol,*}}$, we integrated the convolution of the template spectrum with the blackbody spectral energy distribution at $T_\mathrm{eff}$ over wavelength to derive the flux per stellar surface area, and then integrated over the visible stellar hemisphere with radius $R_*$. To arrive at $L_{\mathrm{Kp,*}}$, we included the Kepler response function~\citep{vancleve2016} in the convolution. Uncertainties on $L_{\mathrm{Kp,*}}$ ranged from 9\;\% to 52\;\% with a median value of 17\;\%.
\section{Methods}
\label{sec:methods}
We developed the open source software \texttt{AltaiPony}\footnote{\url{altaipony.readthedocs.io}} as a toolbox for statistical flare studies tailored but not limited to time series observations obtained by Kepler, K2, and TESS. Its functionality includes light curve de-trending, flare finding and characterization, injection- and recovery of synthetic flares, visualization commands, and methods for the statistical analysis of flare samples. We used \texttt{AltaiPony} to detect flare candidates~(Section~\ref{sec:sec:flarefinding}), and determine their equivalent durations~(Section~\ref{sec:ed}). We validated all events by eye, and calculated their energies in the Kepler band~(Section~\ref{sec:en}). The flare rates are believed to follow a power law distribution that spans multiple orders of magnitude in energy. We used \texttt{AltaiPony}'s \texttt{FFD.fit\_powerlaw()} method to fit the power law parameters simultaneously using the Markov Chain Monte Carlo (MCMC) method to sample from the posterior distribution~(Section~\ref{powerlawfits}).
\subsection{Flare finding}
\label{sec:sec:flarefinding}
We used the open source software \texttt{AltaiPony} to automatically detect and characterize flares in our sample that relied on \texttt{K2SC}\footnote{\url{github.com/OxES/k2sc}}~\citep{aigrain_k2sc_2016} to remove instrumental and astrophysical variablity from K2 light curves. We did not use the de-trended light curves available in MAST\footnote{\url{archive.stsci.edu/prepds/k2sc/}}. Instead, we used \texttt{K2SC} to de-trend light curves from the high quality, uniformly processed and documented final K2 data release ourselves. We clipped outliers at $3\sigma$ iteratively, as compared to the original work, where outliers were clipped at $5\sigma$~\citep{aigrain_k2sc_2016}. We also masked $\sqrt{n}$ data points before and after the sequences of outliers with lengths $n>1$, because these regions often showed physically elevated flux levels below the $3\sigma$ threshold.
\\
After de-trending, the flare finder algorithm searched for continuous observing periods, defined as time series that were longer than 10 data points at a minimum cadence of 2\;h. All further routines were run on these observing periods. We estimated the scatter in the de-trended flux using a rolling standard deviation with a $7.5$ h window excluding the outliers. 
\\
In the next step, the finder iteratively clipped excursions from the median value at $3\sigma$. After each iteration, outliers were replaced with the last median value. Either after convergence, or 50 iterations, the resulting median value was adopted. 
\\
The previously masked outliers in the light curve were assigned the mean uncertainty from the remainder of the light curve. Thus we used the non-flaring parts of the light curve to define the scatter in the flaring parts. This procedure assumed that the scatter of the 80 day light curve was stable throughout the observation. In K2, stability could not be guaranteed, leading to false positive detections in regions of light curves where the uncertainty was underestimated due to strong systematic effects that \texttt{K2SC} could not compensate. These false positives were later removed manually. 
\\
Using the iteratively derived median as quiescent flux and the estimated scatter, we flagged all outliers above $3\sigma$ in the residual light curve. For flare candidates, we required at least three consecutive data points to fulfill the $3\sigma$-criterion. We merged the candidates into single candidate events if they were less than four data points apart. 
\\
With the automated procedure we detected 6916flare candidates. However, the Kepler flare sample has shown to be difficult to treat in a fully automated way. Without manual vetting, the event samples remained significantly contaminated~\citep{yang2019}. As K2 was subject to severe technical difficulties, we expected that the contamination would be even higher. Some light curves could not be de-trended using \texttt{K2SC} alone. Light curves with extreme astrophysical signal like deep transits, rotational modulation on time scales of a few hours or passages of bright solar system objects (SSO) had to be masked or fitted with an additional sinusoidal component to the \texttt{K2SC}-treated time series. Other SSOs mimicked the fast rise and exponential decay shape of flares, and could only be identified using an extension to the Virtual Observatory (VO) tool sky body tracker (SkyBoT, \citealt{berthier2006, berthier2016}). We identified 144 SSOs using SkyBot and visual inspection with \texttt{lightkurve}~\citep{lightkurve2018}, that is 3.6~\%of all confirmed flare candidates~(see Appendix \ref{app:skybot}). A number of light curves were excluded from the flare search because they saturated the detector~($\sim 0.2\%$ of all light curves), or because the target aperture overlapped with broken pixels~(one case). Some very faint targets and extreme variables could not be searched because the de-trending procedure did not terminate successfully~($\sim 0.1\%$ of all light curves). The online version of the final flare sample (Table~\ref{tab:flares}) includes explanatory flags, and notes on the excluded targets. Eventually, we vetted all candidates manually and confirmed 3844events.
%------------------------------------------
\subsection{Equivalent duration $ED$}
\label{sec:ed}
For each candidate, flaring time, amplitude and equivalent duration ($ED$) were returned.
$ED$ is the area between the light curve and the quiescent flux, that is, the integrated flare flux divided by the median quiescent flux $F_0$ of the star, integrated over the flare duration~\citep{gershberg1972}:
\begin{equation}
\label{eq:ED}
ED=\displaystyle \int \mathrm dt\, \frac{F_{flare}(t)}{F_0}.
\end{equation}
$ED$ is a quantity independent of stellar calibration and distance that is suited to compare flaring activity on stars where these properties are uncertain. $ED$ describes the time during which the non-flaring star releases as much energy as the flare itself. This time can be shorter or longer than the actual flare duration. The uncertainty in $ED$ depends on the light curve noise properties, time sampling, spacecraft roll, and potentially other systematic effects introduced by the de-trending procedure. 
%------------------------------------------
\subsection{Kepler flare energy $E_\mathrm{Kp}$}
\label{sec:en}
Multiband time-resolved observations of active M dwarfs have shown that thermal continuum flux accounts for the majority  of the energy radiated by flares at optical wavelengths~\citep{kowalski2013}.
The effective temperature of this blackbody varies:
While solar flares are relatively cool, with \mbox{$T_\mathrm{eff}\approx5\,000-7\,000 $\;K}~\citep{kleint_solarstellarwlf_2016, kerr_solarstellarwlf_2014, watanabe_solarstellarwlf_2013, namekata_solarstellarwlf_2017}, most stellar events emit in the blue, and exhibit temperatures of about $9\,000-10\,000$\;K~\citep{1992ApJS...78..565H, kretzschmar_sun_2011, davenport_multi-wavelength_2012, shibayama2013}. However, in ultraviolet time resolved spectral observations, one superflare was reported at a blackbody temperature of $15\,500$ K~\citep{loyd2018}, and another could even be fit with a single $40\,000$\;K spectral energy distribution~\citep{froning_40000_2019}.
A dependence of flare temperature on stellar age, or mass, or both, would additionally bias our analysis. For example, at about $6\,200$\;K, the Kepler pass band captures the largest flux fraction, at $10\,000$\;K $72\,\%$, at $40\,000$\,K only $4\%$ of this value is transmitted. 
Given the uncertainties on stellar luminosity in the Kepler band (see Section \ref{sec:lum}) and on equivalent duration, the flare energies
\begin{equation}
E_\mathrm{Kp} = L_\mathrm{Kp,*} \cdot ED
\end{equation}  
will substantially deviate from the true released energy. If these uncertainties did not affect all flares in a similar fashion, the present analysis will have suffered from nonuniform biases that affected the flare frequency distributions on stars of different ages and temperatures, and skewed the flare distributions within each subsample. However, the comparison to other studies (see Section \ref{sec:consistency_other_work}) suggested that our results were mostly consistent with studies based on similar data~(see Fig. \ref{fig:otherwork}, and \citealt{lin2019} therein for an example), albeit using different methods to infer $E_\mathrm{flare}$.  

\subsection{Power law fits}
\label{powerlawfits}
Flare frequency distributions (FFDs) follow power law relations that cover several orders of magnitude, from solar microflares to stellar superflares~(see Fig. 9 in \citealt{shibayama2013}). In the cumulative distribution the frequency $f(>E)$ of flares above a certain energy $E$ is defined as
\begin{equation}
f(>E) = \dfrac{\beta}{\alpha - 1}E^{-\alpha + 1},
\label{eqn:cumdist}
\end{equation}
and, analogously, for $ED$~\citep{gershberg1972}. We used and compared two approaches to fitting $\alpha$ and $\beta$ to the data. The first was a modified maximum likelihood estimator (MMLE) for power law distributions~\citep{maschberger2009}. The second approach was more specifically tailored to flaring activity. Following~\citet{wheatland_flaresbayes_2004}, we used the MCMC method to sample from the predictive distribution of flares from a prototype flare source. This source has two characteristics. Firstly, it produces flare events as a homogeneous Poisson process in time. Secondly, the flare energies are power law distributed. While the MMLE was computationally efficient, and useful to obtain first results~(see Appendix~\ref{sec:app:MMLE}), only the predictive distribution allowed us to fit for $\alpha$ and $\beta$ simultaneously, and determine their uncertainties, so we adopted the latter.
\\
The posterior distribution in \citet{wheatland_flaresbayes_2004} captured both the Poissonian distribution of flare events in time, and their power law distribution in energy, simultaneously. The authors derived this model to be able to predict the flaring rate above a given energy for any active region on the Sun including changes in flaring activity rates as the active region evolves, and also characteristics of the active region itself, such as sunspot classifiers. In our simplification of the model, we assumed that the flare generating process did not change within the observation time span in any star in our sample ($M=M'$ in Eq. 24 in \citealt{wheatland_flaresbayes_2004}). Another assumption was that this process was the same for all stars in the sample ($\Lambda_{MC}=1$ in aforementioned Eq. 24). Under these assumptions, the samples of flares found in the light curves of different stars and light curves obtained during different campaigns could be stacked together. With these simplifications to Eq. 24, we defined the joint posterior distribution for the probability $\epsilon$ that a flare with $ED$ or $E_\mathrm{Kp}$ above some value $S_2$ would occur within a time period $\Delta T$:
\begin{eqnarray}
\label{joint_posterior}
p(\epsilon, \alpha) &=& C \cdot\, (-\ln(1 - \epsilon)^{M})\nonumber\\
                    && \cdot\, (\alpha-1)^M \cdot\, \Gamma(\alpha) \left[\dfrac{(S_2 / S_1)^{M+1}}{\pi} \right]^{\alpha}\nonumber\\
                    && \cdot\, (1-\epsilon)^{(T / \Delta T) \,\cdot\, (S_2 /S_1)^{\alpha-1} -1 }.
\end{eqnarray}
$C$ was the normalization constant, $M$ was the number of events, $T$ the total observation time. $\Gamma$ contained the prior distribution for $\alpha$, and $S_1$ denoted the detection threshold above which all flares were detected. $\pi$ encapsulated the flare energies as
\begin{equation}
    \pi = \displaystyle \prod_{i=1}^M \dfrac{s_i}{S_1},
\end{equation}
where $\{s_1,s_2,...s_m\}$ were the flare energies $E_\mathrm{Kp}$ or $ED$. We chose $\Delta T$ to be equal to the total observation time of the FFD, and $S_2$ was set to be ten times the maximum observed energy in that FFD.
\\
From the posterior distribution of $\epsilon$ we derived $\beta$ by combining Poisson statistics 
\begin{equation}
\epsilon = 1 - \mathrm{e}^{(-f\cdot\Delta T)}
\label{poissonstats}
\end{equation}
and the cumulative power law function in Eq. \ref{eqn:cumdist}:
\begin{equation}
\beta = - \dfrac{\ln(1 - \epsilon)\cdot (\alpha -1)}{\Delta T} \cdot S_2^{\alpha -1}
\label{eqn:epstobeta}
\end{equation}
With a uniform prior for $\alpha$ the results from the MMLE and MCMC sampling from the posterior distribution were consistent within uncertainties~(see Appendix \ref{sec:app:MMLE}). However, the MCMC method allowed us to fit for both $\epsilon$ and $\alpha$ simultaneously, and to use more informative priors. 
\\
The power law exponent determined for flare frequency distributions has consistently been found to be independent of age~\citep{davenport2019}, and spectral type for solar-type and low mass dwarfs~(e.g., \citealt{chang2015, howard2019, lin2019}; see Fig. \ref{fig:powerlaw_literature} for an overview). We chose our prior to reflect this result: Starting from a uniform prior for $\alpha$ and $\epsilon$, we found a Gaussian distribution to be an excellent fit to the posterior distribution for $\alpha$ for the full sample of flares in $E_\mathrm{Kp}$ and $ED$ space. We then used this Gaussian distribution as a prior for the FFDs in individual age and $T_\mathrm{eff}$ bins. 
\subsection{Injection and recovery of synthetic flares}
\label{sec:sec:injrec}
Energies of flares found in Kepler 30 minute cadence light curves are underestimated by about 25\,\% compared to 1 minute cadence~\citep{yang_flaresampling_2018}. Moreover, the low energy end of FFDs is incomplete because not all small flares of a given energy are recovered~\citep{davenport_kepler_2016}. In \citetalias{ilin2019} we injected and recovered synthetic flare signal based on the empirical flare template from~\citet{davenport_kepler_2014} to determine the energy bias, and the recovery probability as a function of energy. We injected flares into the K2 light curves after de-trending them with \texttt{K2SC} to test the performance of the flare finding and characterization algorithms. Here, we performed this procedure on a random selection of light curves from our sample before applying \texttt{K2SC}. The injected flare candidates thus had to survive the \texttt{K2SC} de-trending in addition to flare search prescription. As this procedure was computationally very expensive, we did not repeat the process for all light curves in our sample. We noted that both energy bias and recovery probability were better described as a function of amplitude and duration than of energy alone, and that they varied stongly from light curve to light curve. Therefore, energy bias and recovery probability should be quantified light curve by light curve. While this was not possible here, we recommend using the \texttt{FlareLightCurve.sample\_flare\_recovery()} method and associated visualization functions in \texttt{AltaiPony} with light curves from the original Kepler mission, and TESS, to assess the effects of light curve quality, and de-trending and flare search methods on the detected ensembles of flares\footnote{Examples of usage can be found in the software documentation.}. We mitigated the lack of such an assessment by cutting our FFDs at the incomplete low energy tail, but we have to acknowledge that the reported energies remain underestimated~\citep{yang_flaresampling_2018}. \citet{aschwanden_powerlaws_2015} suggested that the observed incompleteness may be partly canceled by background contamination from, for instance, cosmic rays~, or solar system objects (SSO). While the effect of cosmic rays is currently not well constrained~\citepalias{ilin2019}, we could quantify the contamination by cataloged SSOs to 3.6~\%~(see Section \ref{sec:sec:flarefinding}).
%----------------------------------------
\section{Results}
\label{sec:results}
%----------------------------------------
\begin{table*}
\caption{Confirmed flare candidates detected in open cluster stars observed by K2, sorted by amplitude $a$.}
\label{tab:flares}
\centering
\footnotesize
\begin{tabular}{lccccccccccr}
\hline\hline
      EPIC &   C &   cluster &   $c_0$ &   $c_1$ &        $a$ & $T_\mathrm{eff}$ [K] &    $ED$ [s] & $L_\mathrm{bol,*}$ [erg/s] & $L_\mathrm{Kp}$ [erg/s] \\
\hline
 211079830 &   4 &  Pleiades &  105984 &  105993 &  15.330194 &             3097(87) &   63340(63) &        $8.29(3.34)10^{29}$ &     $3.32(1.34)10^{29}$ \\
 210720772 &   4 &  Pleiades &  107181 &  107184 &   9.674757 &             3104(86) &    19901(7) &        $8.66(3.45)10^{29}$ &     $3.48(1.38)10^{29}$ \\
 247523445 &  13 &    Hyades &  143106 &  143109 &   8.260956 &             2964(49) &   16615(11) &        $4.33(1.09)10^{29}$ &     $1.69(0.43)10^{29}$ \\
 212021131 &   5 &  Praesepe &  108974 &  108980 &   7.421916 &             3215(68) &  19828(175) &        $2.01(0.65)10^{30}$ &     $8.22(2.65)10^{29}$ \\
 210978953 &   4 &  Pleiades &  106762 &  106770 &   6.769888 &             3050(95) &  39467(125) &        $6.20(2.63)10^{29}$ &     $2.46(1.04)10^{29}$ \\
 211913613 &  16 &  Praesepe &  156845 &  156849 &   6.690356 &             3218(66) &  21632(147) &        $2.05(0.65)10^{30}$ &     $8.37(2.63)10^{29}$ \\
 211127297 &   4 &  Pleiades &  106754 &  106759 &   6.449569 &             3147(86) &    20830(5) &        $1.13(0.45)10^{30}$ &     $4.59(1.81)10^{29}$ \\
 211681193 &   5 &  Praesepe &  108116 &  108120 &   5.570804 &             3182(76) &   11198(95) &        $1.64(0.58)10^{30}$ &     $6.68(2.37)10^{29}$ \\
 211024798 &   4 &  Pleiades &  104822 &  104826 &   5.394897 &             3290(62) &   12524(63) &        $2.58(0.75)10^{30}$ &     $1.06(0.31)10^{30}$ \\
 211134185 &   4 &  Pleiades &  103891 &  103896 &   4.951629 &             3127(96) &   17459(33) &        $9.99(4.33)10^{29}$ &     $4.03(1.75)10^{29}$ \\
 211095280 &   4 &  Pleiades &  106283 &  106287 &   4.796979 &             3138(93) &   12186(10) &        $1.06(0.45)10^{30}$ &     $4.32(1.82)10^{29}$ \\
 211022535 &   4 &  Pleiades &  104262 &  104267 &   3.994973 &             2953(76) &   15164(57) &        $3.50(1.10)10^{29}$ &     $1.37(0.43)10^{29}$ \\
 211010517 &   4 &  Pleiades &  106680 &  106685 &   3.968907 &             3252(70) &  16241(171) &        $2.09(0.68)10^{30}$ &     $8.59(2.77)10^{29}$ \\
 210846442 &   4 &  Pleiades &  104410 &  104415 &   3.959977 &             3311(79) &   13515(10) &        $2.88(0.96)10^{30}$ &     $1.18(0.40)10^{30}$ \\
 212017838 &   5 &  Praesepe &  111183 &  111192 &   3.671037 &             3307(89) &   10274(10) &        $3.35(1.22)10^{30}$ &     $1.38(0.50)10^{30}$ \\
 211984058 &   5 &  Praesepe &  109952 &  109965 &   3.320636 &             3124(97) &  16063(149) &        $1.17(0.51)10^{30}$ &     $4.73(2.06)10^{29}$ \\
 211912899 &   5 &  Praesepe &  110700 &  110706 &   3.257997 &             3133(81) &    7967(16) &        $1.24(0.47)10^{30}$ &     $5.01(1.89)10^{29}$ \\
 211151674 &   4 &  Pleiades &  106457 &  106467 &   3.166323 &             3072(93) &  17843(106) &        $7.10(2.99)10^{29}$ &     $2.84(1.19)10^{29}$ \\
 211822895 &   5 &  Praesepe &  107809 &  107812 &   3.164494 &            3005(177) &    6883(31) &        $5.68(3.98)10^{29}$ &     $2.25(1.57)10^{29}$ \\
 211760567 &   5 &  Praesepe &  109962 &  109965 &   3.061417 &             3245(58) &     7838(9) &        $2.39(0.69)10^{30}$ &     $9.84(2.83)10^{29}$ \\
 211939350 &  16 &  Praesepe &  155018 &  155022 &   2.892939 &            3149(150) &    9542(68) &        $1.36(0.86)10^{30}$ &     $5.52(3.49)10^{29}$ \\
 211137806 &   4 &    Hyades &  106768 &  106775 &   2.482828 &             3127(58) &     9208(8) &        $1.16(0.35)10^{30}$ &     $4.69(1.40)10^{29}$ \\
 210674207 &  13 &    Hyades &  141708 &  141711 &   2.419586 &             3210(74) &     5343(3) &        $1.90(0.65)10^{30}$ &     $7.78(2.66)10^{29}$ \\
 211994910 &   5 &  Praesepe &  109858 &  109864 &   2.401426 &             3325(95) &    7807(14) &        $3.71(1.40)10^{30}$ &     $1.53(0.58)10^{30}$ \\
 211010517 &   4 &  Pleiades &  106561 &  106564 &   2.341076 &             3252(70) &    4838(36) &        $2.09(0.68)10^{30}$ &     $8.59(2.77)10^{29}$ \\
 211984058 &   5 &  Praesepe &  108584 &  108587 &   2.155468 &             3124(97) &    4817(13) &        $1.17(0.51)10^{30}$ &     $4.73(2.06)10^{29}$ \\
 211983544 &  18 &  Praesepe &  162792 &  162798 &   2.096077 &             3159(98) &    6924(55) &        $1.44(0.63)10^{30}$ &     $5.88(2.55)10^{29}$ \\
 210988354 &   4 &  Pleiades &  106012 &  106017 &   2.057812 &            3661(144) &     5725(1) &        $1.40(0.31)10^{31}$ &     $6.06(1.32)10^{30}$ \\
 210886447 &   4 &  Pleiades &  103871 &  103877 &   2.057490 &            3048(105) &    8710(49) &        $6.13(2.83)10^{29}$ &     $2.43(1.12)10^{29}$ \\
 211098921 &   4 &  Pleiades &  105254 &  105262 &   2.041041 &            3300(101) &     7499(5) &        $2.72(1.10)10^{30}$ &     $1.12(0.45)10^{30}$ \\
\hline

\end{tabular}

\tablefoot{The full table is available via CDS. EPIC and C denote the K2 ID and observing campaign. $a$ $-$ relative flare amplitude, $c_0$ and $c_1$ $-$ flare start and end cadence numbers, $T_\mathrm{eff}$ $-$ stellar effective temperature, $ED$ $-$ equivalent duration,  $L_\mathrm{bol,*}$ $-$ projected stellar bolometric luminosity, $L_\mathrm{Kp}$ - projected stellar luminosity in the Kepler band. Uncertainties are given in parentheses.}
\end{table*}
%----------------------------------------

The core objective of this work was to quantify how the previously noted decline in flaring activity with age would unfold for different spectral types. To this end, we searched the long cadence light curves of stars across a broad range of spectral types in five different open clusters for flares, and measured their energies and occurrence rates. We found 3844\unskip\;flares on 2111\unskip\;stars in five open clusters for $T_\mathrm{eff}$ between 2500 K and 6000 K. The flares' amplitudes $a$, start and end cadence numbers $c_0$ and $c_1$, $ED$, energies $E_\mathrm{Kp}$, along with the stars' EPIC ID, $T_\mathrm{eff}$, and projected bolometric and Kepler band luminosities $L_\mathrm{bol,*}$ and $L_\mathrm{Kp,*}$ are listed in Table \ref{tab:flares}. We fitted power law relations to the flare frequency distributions (FFD) of stars binned by age, and $T_\mathrm{eff}$~(Section~\ref{sec:sec:ffds}, Figs. \ref{fig:powerlawfits_s}, \ref{fig:powerlawfits_erg} and Table \ref{tab:powerlawtable_spt}). We found that flaring activity decreased with increasing age from ZAMS to 3.6 Gyr, and from mid-M to G dwarfs~(Sections~\ref{sec:beta100s} and~\ref{sec:fa}, Figs.~\ref{fig:beta_T_age} and~\ref{fig:FA}). Except for the stars in the coolest temperature bin (M5.5-M8, 2\,500-3\,000 K), stellar flaring activity at a given age was higher for cooler stars. The stars in our sample indicated a mass and rotation dependent threshold age above which they were no longer found on the saturated activity branch which was characterized by fast rotating and actively flaring stars. Stars above this threshold showed much lower flaring rates~(Sections~\ref{sec:hyaprarot}~and~\ref{sec:4dim}, Fig.~\ref{fig:4dim}). In the old clusters Ruprecht 147 and M67 we found very few flares, and suggest that a significant fraction of those found on higher mass stars may stem from unresovled low-mass companions~(Section~\ref{sec:m67r147}). 
%------------------------------------------------
\subsection{Flare frequency distributions}
\label{sec:sec:ffds}
\begin{figure*}[ht!]
    \centering
    \includegraphics[width=14cm]{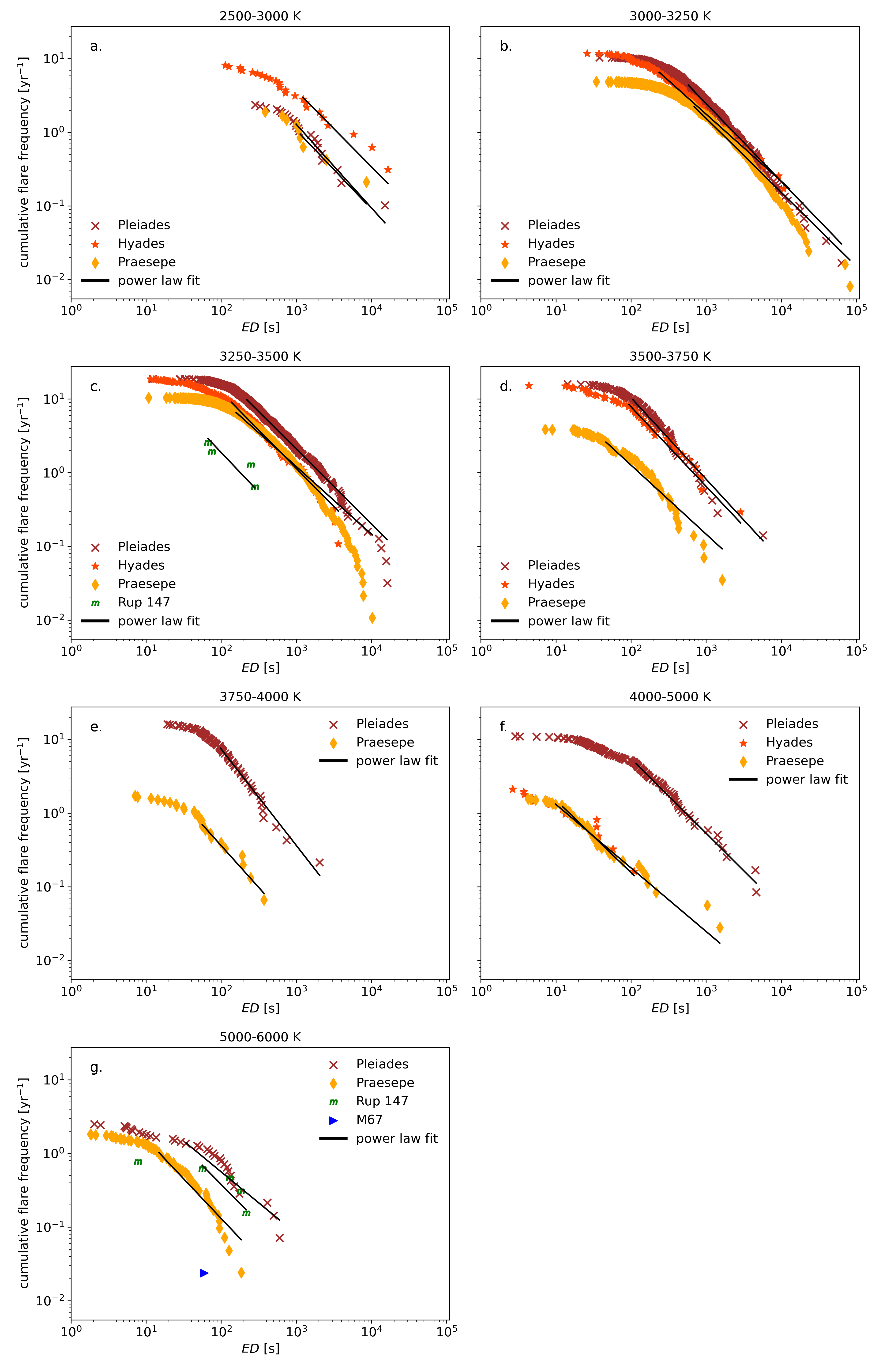}
    \caption{Cumulative flare frequency distributions (FFD, scatter) in $ED$, and respective power law fits to the portion of the FFDs where we considered our sample not be affected by reduced detection efficiency of low-energy flares (black lines), in log-log representation. In each of the panels a.-g. the distributions of flares are shown for different open clusters for samples of stars in the $T_\mathrm{eff}$ bin that is given in the panel title. The different clusters are the Pleiades (maroon crosses), the Hyades (orange stars), Praesepe (yellow diamonds), Ruprecht 147 (green "m"s), and M67 (blue triangles). The extension of the black lines in $ED$ indicates the low-energy threshold adopted for every fit. When no distribution is shown, either no flares were found in the sub-sample, or no stars in that $T_\mathrm{eff}$ bin were observed by K2. The overview of all power law fit parameters including uncertainties is given in Table \ref{tab:powerlawtable_spt}.}       	
    \label{fig:powerlawfits_s}
\end{figure*}

\begin{figure*}[ht!]
    \centering
    \includegraphics[width=14.5cm]{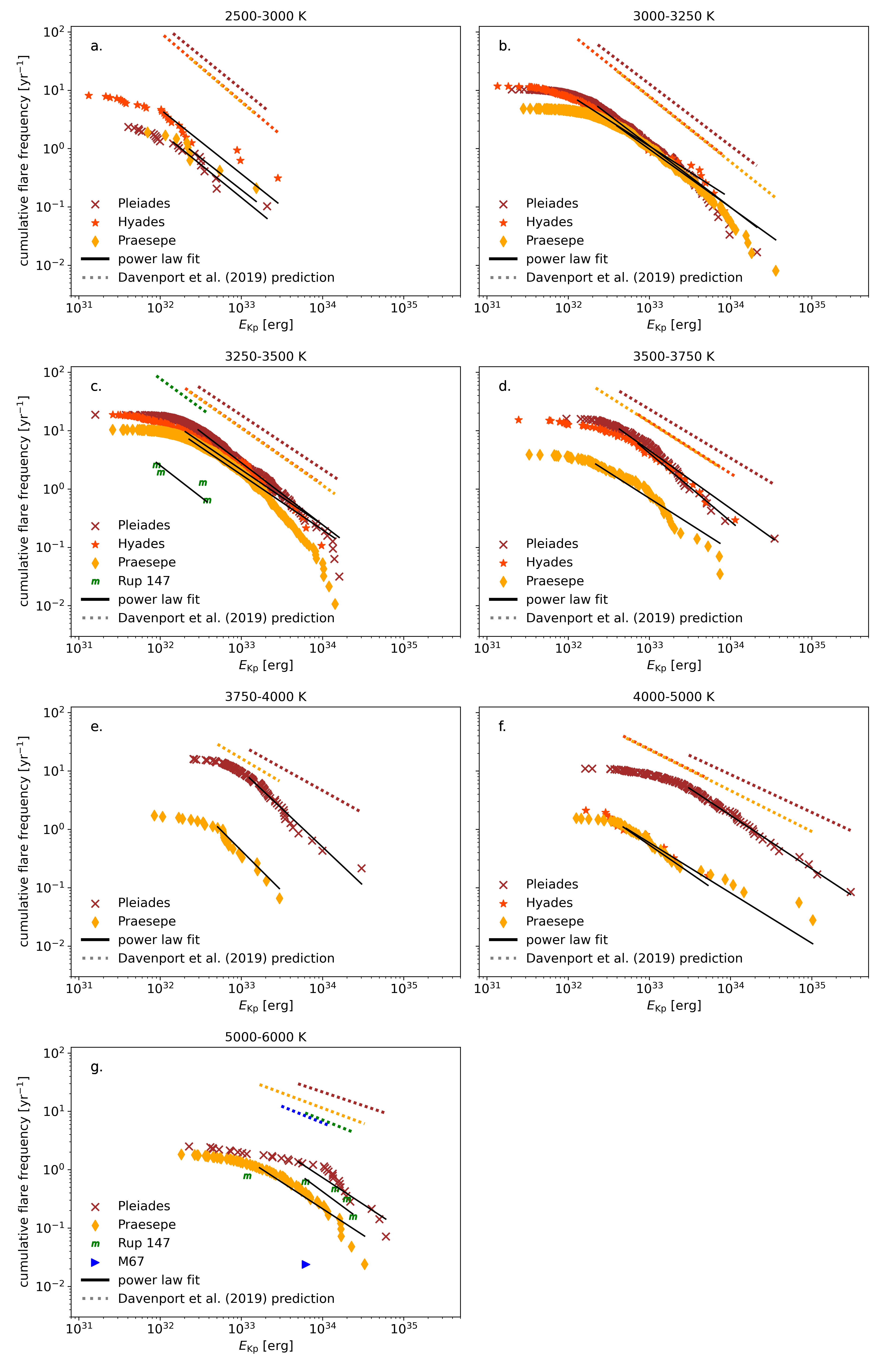}
    \caption{Same as in Fig. \ref{fig:powerlawfits_s} but for $E_\mathrm{Kp}$ instead of $ED$. Additionally, predictions from the \citet{davenport2019} gyrochronology-calibrated FFD parametrization with stellar age and mass were added in the colors that denote the different open clusters~(see discussion in Section \ref{sec:davenport}). To calculate the predictions, we used the mean of each $T_\mathrm{eff}$ bin to determine the mass~\citep{pecaut_intrinsic_2013}, and open cluster ages from the literature.}        	
    \label{fig:powerlawfits_erg}
\end{figure*}

\begin{figure}[ht!]
    \centering
    \includegraphics[width=\hsize]{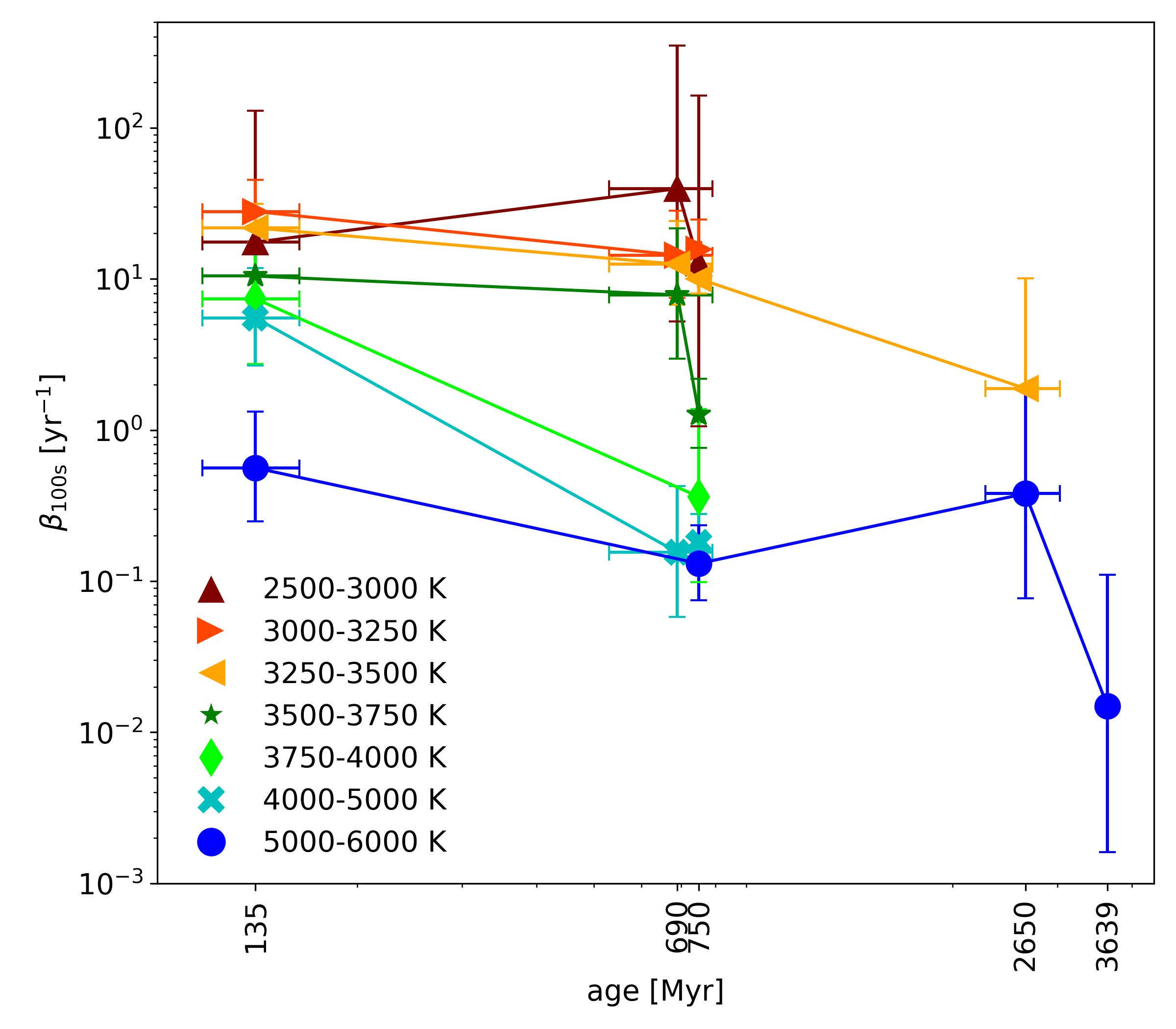}
    \caption{The power law fit intercept $\beta_\mathrm{s}$ for the flare frequency distributions in Fig.~\ref{fig:powerlawfits_s} as a function of age. The values are grouped by $T_\mathrm{eff}$ (different colors and symbols in the legend), and presented on a log-log scale. $\beta_\mathrm{s}$ indicates the occurrence rate of flares with $ED=1$ s~(see Eq. \ref{eqn:cumdist}). It is controlled for stellar luminosity, so that we can compare flaring activity across both age, and $T_\mathrm{eff}$, at the same time. The age labels correspond to the open cluster ages in Table \ref{tab:data_clusters}.}    	
    \label{fig:beta_T_age}
\end{figure}

\begin{figure}[ht!]
    \centering
    \includegraphics[width=\hsize]{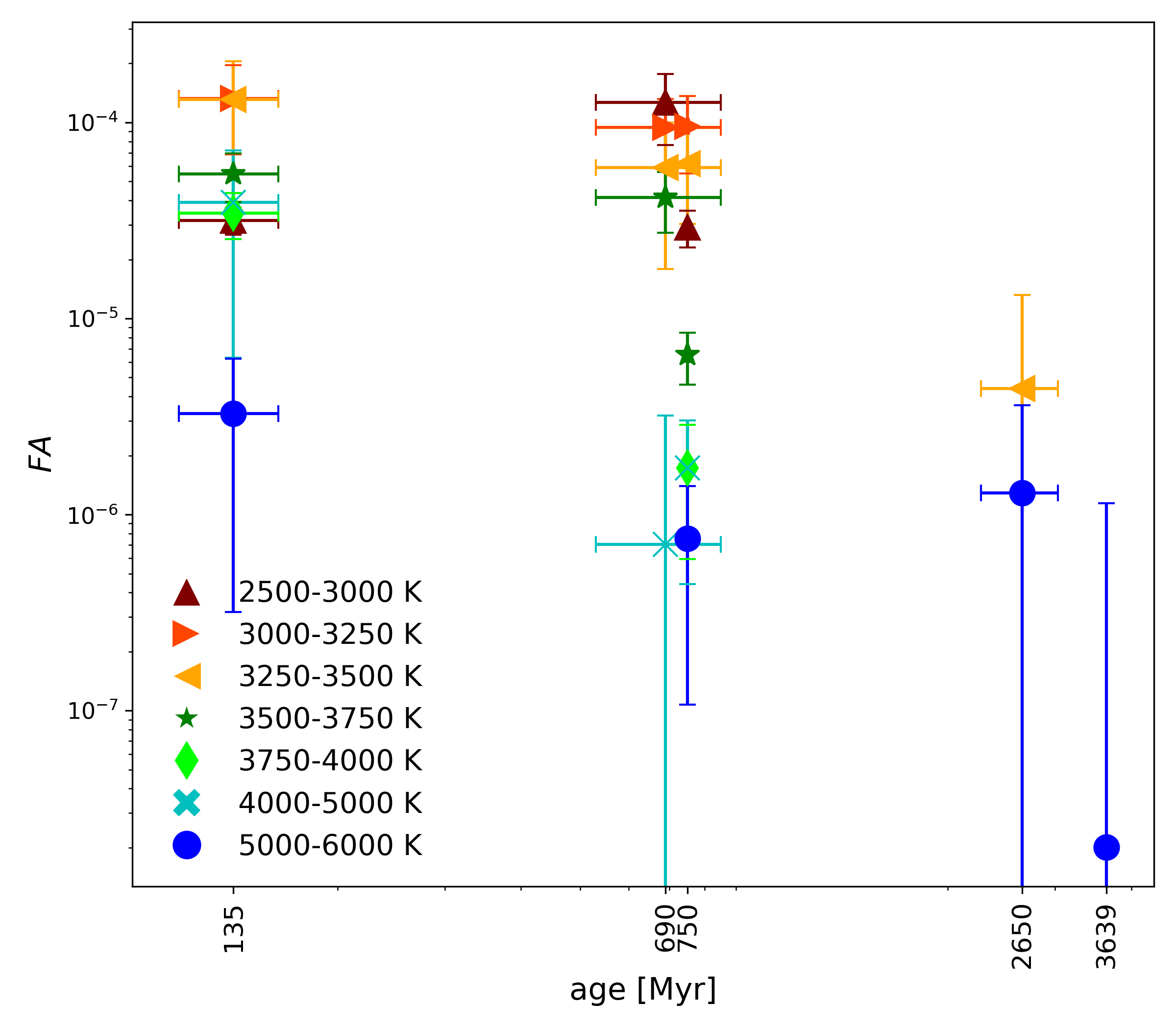}
    \caption{Median fraction of stellar luminosity emitted in flares ($FA$), and standard deviation plotted against age. The values are grouped by $T_\mathrm{eff}$, and presented on a log-log scale. The colors and shapes are the same as in Fig.~\ref{fig:beta_T_age}. We calculated the shared energy detection threshold $3.31\cdot 10^{32}$ erg using the method described in Section \ref{sec:sec:ffds} to derive $FA$ with Eq.~\ref{eq:FA}. $FA$ is a useful measure of relative stellar activity as long as the flux portion of the quiescent star in the Kepler band is roughly constant. It is therefore more meaningful to compare $FA$ across age than across $T_\mathrm{eff}$.}   
    \label{fig:FA}
\end{figure}

The aim of this work was to investigate the effects of age on the flaring activity for low mass stars. We therefore split up the full sample into $T_\mathrm{eff}$ bins and constructed the FFDs cluster by cluster~(Figs. \ref{fig:powerlawfits_s} and \ref{fig:powerlawfits_erg}). We used a broad Gaussian prior for $\alpha$ with mean 1.9 and standard deviation .4 that covered the range of $\alpha$ values obtained in past studies~(see Fig.~\ref{fig:powerlaw_literature} in the discussion). The power law fit parameters to these FFDs, broken down by cluster and $T_\mathrm{eff}$, are summarized in Table~\ref{tab:powerlawtable_spt}. In the table, $\alpha$ and $\beta$ are the power law parameters from Eq.~\ref{eqn:cumdist}. The number of flares $n_\mathrm{tot}$ in the distribution is higher than the number $n_\mathrm{fit}$ of flares used to fit the power law because we omitted the low-energy tail of the distribution where the detection probability decreased. Our approximation to the detection threshold was the highest minimum detected energy
from all FFDs of individual stars in a given bin. This was equivalent to using only the part of the FFD that was above the detection threshold for all stars in that bin. The indices s and erg to $\alpha$ and $\beta$ indicate the fits to the FFDs in $ED$~(Fig.~\ref{fig:powerlawfits_s}) and $E_\mathrm{Kp}$~(Fig.~\ref{fig:powerlawfits_erg}), respectively. For the single flare detection in M67 we give the power law parameters obtained directly from Eq.~\ref{eqn:cumdist} using the prior value and uncertainty for $\alpha$. We provide Table \ref{tab:powerlawtable_spt} to enable future science to build a comprehensive flare rate evolution model.
%-------------------------------------------

%----------------------------------------
%----------------------------------------
%----------------------------------------
%----------------------------------------

%-------------------------------------------
\begin{sidewaystable*}
\caption{Power law fit parameters $\alpha$ and $\beta$ for all clusters and $T_\mathrm{eff}$ bins in $E_\mathrm{Kp}$ and $ED$ FFDs.}
\label{tab:powerlawtable_spt}
\centering
\begin{tabular}{lccccccccccccr}
\hline
          &         &                $\alpha_\mathrm{s}$ &      $\beta_\mathrm{s}$ [yr$^{-1}$] & $ED_\mathrm{min}$ [s] & $ED_\mathrm{max}$ [s] &  $n_\mathrm{tot,s}$ &  $n_\mathrm{fit,s}$ &              $\alpha_\mathrm{erg}$ &                                          $\beta_\mathrm{erg}$ [yr$^{-1}$] & $E_\mathrm{Kp,min}$ [erg] & $E_\mathrm{Kp,max}$ [erg] &  $n_\mathrm{tot,erg}$ &  $n_\mathrm{fit,erg}$ \\
Teff & cluster &                                    &                                     &                       &                       &                     &                     &                                    &                                                                           &                           &                           &                       &                       \\
\hline
2500-3000 & Hyades &  $2.03\left(^{0.27}_{0.24}\right)$ &  $4707\left(^{37065}_{4086}\right)$ &    $1.24\cdot 10^{3}$ &    $1.66\cdot 10^{4}$ &                  26 &                   9 &  $2.11\left(^{0.25}_{0.23}\right)$ &  $1.4 \cdot 10^{36}\left(^{1.5 \cdot 10^{44}}_{1.4 \cdot 10^{36}}\right)$ &       $1.11\cdot 10^{32}$ &       $2.82\cdot 10^{33}$ &                    26 &                    13 \\
          & Pleiades &  $2.13\left(^{0.26}_{0.23}\right)$ &  $3668\left(^{23493}_{3102}\right)$ &    $0.99\cdot 10^{3}$ &    $1.52\cdot 10^{4}$ &                  23 &                  12 &  $2.13\left(^{0.26}_{0.24}\right)$ &  $3.3 \cdot 10^{36}\left(^{6.6 \cdot 10^{44}}_{3.3 \cdot 10^{36}}\right)$ &       $1.45\cdot 10^{32}$ &       $2.08\cdot 10^{33}$ &                    23 &                    12 \\
          & Praesepe &  $2.07\left(^{0.32}_{0.30}\right)$ &  $1883\left(^{22416}_{1727}\right)$ &    $1.13\cdot 10^{3}$ &    $0.86\cdot 10^{4}$ &                   9 &                   4 &  $2.09\left(^{0.32}_{0.30}\right)$ &  $2.1 \cdot 10^{35}\left(^{8.8 \cdot 10^{45}}_{2.1 \cdot 10^{35}}\right)$ &        $2.3\cdot 10^{32}$ &       $1.53\cdot 10^{33}$ &                     9 &                     4 \\
3000-3250 & Hyades &  $1.91\left(^{0.10}_{0.10}\right)$ &      $865\left(^{832}_{413}\right)$ &    $2.37\cdot 10^{2}$ &    $1.29\cdot 10^{4}$ &                 137 &                  76 &  $1.89\left(^{0.10}_{0.09}\right)$ &  $2.3 \cdot 10^{29}\left(^{3.8 \cdot 10^{32}}_{2.3 \cdot 10^{29}}\right)$ &       $1.31\cdot 10^{32}$ &       $0.84\cdot 10^{34}$ &                   137 &                    78 \\
          & Pleiades &  $2.06\left(^{0.06}_{0.06}\right)$ &   $3809\left(^{2338}_{1431}\right)$ &    $0.58\cdot 10^{3}$ &    $0.63\cdot 10^{5}$ &                 616 &                 258 &  $2.06\left(^{0.06}_{0.06}\right)$ &    $8.1 \cdot 10^{34}\left(^{7.4 \cdot 10^{36}}_{8.0 \cdot 10^{34}}\right)$ &       $2.32\cdot 10^{32}$ &       $2.11\cdot 10^{34}$ &                   616 &                   311 \\
          & Praesepe &  $2.00\left(^{0.06}_{0.06}\right)$ &     $1605\left(^{935}_{573}\right)$ &     $0.7\cdot 10^{3}$ &    $0.82\cdot 10^{5}$ &                 601 &                 275 &  $2.00\left(^{0.06}_{0.05}\right)$ &  $8.8 \cdot 10^{32}\left(^{6.9 \cdot 10^{34}}_{8.6 \cdot 10^{32}}\right)$ &        $0.4\cdot 10^{33}$ &       $0.36\cdot 10^{35}$ &                   601 &                   302 \\
3250-3500 & Hyades &  $2.04\left(^{0.11}_{0.10}\right)$ &    $1545\left(^{1434}_{714}\right)$ &    $1.37\cdot 10^{2}$ &    $0.36\cdot 10^{4}$ &                 174 &                  83 &  $1.94\left(^{0.10}_{0.09}\right)$ &  $2.8 \cdot 10^{31}\left(^{4.5 \cdot 10^{34}}_{2.8 \cdot 10^{31}}\right)$ &       $2.04\cdot 10^{32}$ &       $0.97\cdot 10^{34}$ &                   174 &                    90 \\
          & Pleiades &  $2.02\left(^{0.06}_{0.06}\right)$ &    $2367\left(^{1063}_{713}\right)$ &    $2.19\cdot 10^{2}$ &    $1.62\cdot 10^{4}$ &                 590 &                 310 &  $2.06\left(^{0.06}_{0.06}\right)$ &  $3.3 \cdot 10^{35}\left(^{2.9 \cdot 10^{37}}_{3.3 \cdot 10^{35}}\right)$ &       $2.95\cdot 10^{32}$ &       $1.61\cdot 10^{34}$ &                   590 &                   327 \\
          & Praesepe &  $1.92\left(^{0.04}_{0.04}\right)$ &      $626\left(^{164}_{127}\right)$ &    $1.59\cdot 10^{2}$ &    $1.03\cdot 10^{4}$ &                 970 &                 612 &  $1.95\left(^{0.04}_{0.04}\right)$ &  $2.6 \cdot 10^{31}\left(^{3.8 \cdot 10^{32}}_{2.4 \cdot 10^{31}}\right)$ &       $2.28\cdot 10^{32}$ &       $1.42\cdot 10^{34}$ &                   970 &                   663 \\
          & Rup 147 &  $2.07\left(^{0.32}_{0.30}\right)$ &     $283\left(^{1232}_{228}\right)$ &    $0.67\cdot 10^{2}$ &    $2.81\cdot 10^{2}$ &                   4 &                   4 &  $2.08\left(^{0.33}_{0.30}\right)$ &          $1.0\cdot 10^{35}\left(^{3.6 \cdot 10^{45}}_{1.0\cdot 10^{35}}\right)$ &        $0.9\cdot 10^{32}$ &       $0.38\cdot 10^{33}$ &                     4 &                     4 \\
3500-3750 & Hyades &  $2.08\left(^{0.19}_{0.17}\right)$ &    $1216\left(^{2144}_{754}\right)$ &    $0.93\cdot 10^{2}$ &    $2.88\cdot 10^{3}$ &                  52 &                  28 &  $2.17\left(^{0.22}_{0.21}\right)$ &  $1.9 \cdot 10^{39}\left(^{5.6 \cdot 10^{46}}_{1.9 \cdot 10^{39}}\right)$ &       $0.72\cdot 10^{33}$ &       $1.14\cdot 10^{34}$ &                    52 &                    20 \\
          & Pleiades &  $2.11\left(^{0.13}_{0.12}\right)$ &    $1882\left(^{2020}_{943}\right)$ &    $1.06\cdot 10^{2}$ &    $0.57\cdot 10^{4}$ &                 113 &                  69 &  $1.99\left(^{0.11}_{0.11}\right)$ &  $2.7 \cdot 10^{33}\left(^{1.6 \cdot 10^{37}}_{2.7 \cdot 10^{33}}\right)$ &       $0.43\cdot 10^{33}$ &       $0.35\cdot 10^{35}$ &                   113 &                    75 \\
          & Praesepe &  $1.94\left(^{0.11}_{0.10}\right)$ &         $88\left(^{63}_{35}\right)$ &    $0.46\cdot 10^{2}$ &    $1.63\cdot 10^{3}$ &                 111 &                  74 &  $1.89\left(^{0.10}_{0.09}\right)$ &  $1.1 \cdot 10^{29}\left(^{2.2 \cdot 10^{32}}_{1.1 \cdot 10^{29}}\right)$ &       $2.19\cdot 10^{32}$ &       $0.74\cdot 10^{34}$ &                   111 &                    76 \\
3750-4000 & Pleiades &  $2.31\left(^{0.20}_{0.18}\right)$ &   $3990\left(^{7449}_{2512}\right)$ &    $0.98\cdot 10^{2}$ &    $2.05\cdot 10^{3}$ &                  75 &                  35 &  $2.32\left(^{0.20}_{0.18}\right)$ &  $3.4 \cdot 10^{44}\left(^{1.2 \cdot 10^{51}}_{3.4 \cdot 10^{44}}\right)$ &       $1.25\cdot 10^{33}$ &       $3.03\cdot 10^{34}$ &                    75 &                    36 \\
          & Praesepe &  $2.13\left(^{0.27}_{0.25}\right)$ &        $76\left(^{213}_{55}\right)$ &    $0.56\cdot 10^{2}$ &    $0.37\cdot 10^{3}$ &                  26 &                  10 &  $2.39\left(^{0.27}_{0.24}\right)$ &  $4.4 \cdot 10^{45}\left(^{3.4 \cdot 10^{54}}_{4.4 \cdot 10^{45}}\right)$ &       $0.51\cdot 10^{33}$ &       $2.95\cdot 10^{33}$ &                    26 &                    16 \\
4000-5000 & Hyades &  $1.98\left(^{0.28}_{0.25}\right)$ &          $14\left(^{24}_{8}\right)$ &    $1.22\cdot 10^{1}$ &     $1.1\cdot 10^{2}$ &                  13 &                   7 &  $1.96\left(^{0.29}_{0.26}\right)$ &      $2.0 \cdot 10^{31}\left(^{6.1 \cdot 10^{40}}_{2.0 \cdot 10^{31}}\right)$ &       $0.48\cdot 10^{33}$ &       $0.53\cdot 10^{34}$ &                    13 &                     6 \\
          & Pleiades &  $2.02\left(^{0.13}_{0.12}\right)$ &      $603\left(^{681}_{312}\right)$ &    $1.17\cdot 10^{2}$ &    $0.46\cdot 10^{4}$ &                 131 &                  55 &  $1.92\left(^{0.12}_{0.11}\right)$ &  $2.3 \cdot 10^{31}\left(^{2.4 \cdot 10^{35}}_{2.3 \cdot 10^{31}}\right)$ &       $3.06\cdot 10^{33}$ &       $3.01\cdot 10^{35}$ &                   131 &                    59 \\
          & Praesepe &  $1.86\left(^{0.12}_{0.11}\right)$ &            $8.2\left(^{4.5}_{2.8}\right)$ &    $0.99\cdot 10^{1}$ &    $1.52\cdot 10^{3}$ &                  56 &                  47 &  $1.86\left(^{0.14}_{0.12}\right)$ &  $1.3 \cdot 10^{28}\left(^{4.1 \cdot 10^{32}}_{1.3 \cdot 10^{28}}\right)$ &       $0.52\cdot 10^{33}$ &       $1.03\cdot 10^{35}$ &                    56 &                    37 \\
5000-6000 & M67 &  $1.90\left(^{0.40}_{0.40}\right)$ &            $0.9\left(^{5.4}_{0.8}\right)$ &     $0.6\cdot 10^{2}$ &     $0.6\cdot 10^{2}$ &                   1 &                   1 &  $1.90\left(^{0.40}_{0.40}\right)$ &  $5.6 \cdot 10^{28}\left(^{2.7 \cdot 10^{42}}_{5.6 \cdot 10^{28}}\right)$ &       $0.62\cdot 10^{34}$ &       $0.62\cdot 10^{34}$ &                     1 &                     1 \\
          & Pleiades &  $1.84\left(^{0.18}_{0.16}\right)$ &         $22\left(^{30}_{12}\right)$ &    $0.34\cdot 10^{2}$ &     $0.6\cdot 10^{3}$ &                  35 &                  19 &  $1.91\left(^{0.19}_{0.17}\right)$ &    $6.9 \cdot 10^{30}\left(^{2 \cdot 10^{37}}_{6.9 \cdot 10^{30}}\right)$ &        $0.5\cdot 10^{34}$ &        $0.6\cdot 10^{35}$ &                    35 &                    19 \\
          & Praesepe &  $2.08\left(^{0.16}_{0.14}\right)$ &          $20\left(^{15}_{8}\right)$ &    $1.49\cdot 10^{1}$ &    $1.85\cdot 10^{2}$ &                  76 &                  42 &  $1.91\left(^{0.13}_{0.12}\right)$ &  $1.2 \cdot 10^{30}\left(^{3.2 \cdot 10^{34}}_{1.2 \cdot 10^{30}}\right)$ &       $1.68\cdot 10^{33}$ &       $0.33\cdot 10^{35}$ &                    76 &                    44 \\
          & Rup 147 &  $2.03\left(^{0.32}_{0.29}\right)$ &        $45\left(^{182}_{36}\right)$ &    $0.56\cdot 10^{2}$ &    $2.16\cdot 10^{2}$ &                   5 &                   4 &  $2.03\left(^{0.32}_{0.29}\right)$ &  $4.3 \cdot 10^{34}\left(^{3.6 \cdot 10^{45}}_{4.3 \cdot 10^{34}}\right)$ &       $0.61\cdot 10^{34}$ &       $2.36\cdot 10^{34}$ &                     5 &                     4 \\
\hline

\end{tabular}

\tablefoot{Uncertainties are given in parentheses. Indices $_\mathrm{s}$ and $_\mathrm{erg}$ refer to the units of $ED$ and $E_\mathrm{Kp}$ FFDs. $E_\mathrm{Kp, min}$/$ED_\mathrm{min}$ and $E_\mathrm{Kp, max}$/$ED_\mathrm{max}$ are minimum and maximum detected flare $ED$/energy, $n_\mathrm{tot}$ is the total number of flares in the sample. $n_\mathrm{fit}$ is the number of flares used for the power law fit.}
\end{sidewaystable*}
%-------------------------------------------

%-------------------------------------------
%-------------------------------------------
%-------------------------------------------
%---------------------------------------------
\subsection{Flaring activity measure $\beta_\mathrm{100s}$}
\label{sec:beta100s}
Relative flaring activity levels are best described by $\beta_s$, that is, the power law fit intercept $\beta$~(see Eq.~\ref{eqn:cumdist}) for the FFDs in $ED$ space (equivalent to $R_\mathrm{1s}$ in \citealt{davenport2019}). Because of the definition of $ED$ as relative to the quiescent flux of the star (Eq. \ref{eq:ED}), $\beta_s$ is a measure of flaring activity in which the different energy budgets of stars are controlled for, in contrast to $\beta_\mathrm{erg}$, which uses the absolute flare energies. However $\beta_s$ is defined as the flaring rate at $ED>1$\,s which is below the detection threshold for most $T_\mathrm{eff}$-age bins. To reduce errors arising from extrapolation we chose $\beta_\mathrm{100s}$~(analogous to $R_\mathrm{100s}$), that is, Eq.~\ref{eqn:cumdist} evaluated at $ED=100\,$s where most FFDs overlap to compare across both $T_\mathrm{eff}$ and age at the same time.
\\
Fig. \ref{fig:beta_T_age} can be interpreted as a synthesis of the FFDs presented in Fig. \ref{fig:powerlawfits_s}. Overall, $\beta_\mathrm{100s}$ tended to be lower for hotter stars at a given age. This general trend is consistent with the flaring rates found in nearby young moving groups, open clusters, OB associations, and star forming regions with ages between 1 and 800 Myr in TESS~\citep{feinstein2020}. $\beta_\mathrm{100s}$ declined with age in all investigated $T_\mathrm{eff}$ bins, and it declined stronger in the hotter stars, a trend also observed in the transition from 125-300 Myr to 300-800 Myr in the TESS study. An exception to this rule is the $2500-3000$ K bin, where $\beta_\mathrm{100s}$ marginally increased from ZAMS to Hyades age. A slight increase in activity is also seen for stars $<4000\,$K at ages between 1-20 Myr and 30-45 Myr in~\cite{feinstein2020}. This uptick in activity was also present in comparing fast and slow rotators in a K2 short cadence K7-M6 dwarf flaring study~\citep{raetz2020}. In the lowest mass bin (M5-M6) the authors measured higher flaring activity for relatively slow rotators ($P_\mathrm{rot} > 2\,$d) than for their fast counterparts ($P_\mathrm{rot} < 2\,$d).  
\\
\citet{mondrik2019} found evidence of increased flaring activity at intermediate rotation rates between fast and slow sequence~\citep{barnes_rotational_2003}. This finding could not be reproduced in the data obtained from the all-sky EvryFlare survey~\citep{howard2020}. We found $\beta_\mathrm{100s}$ to decrease with age from ZAMS to 700\,Myr for all stars except those with spectral types later than M5~(Fig. \ref{fig:beta_T_age}\,a.). But this does not exclude that $\beta_\mathrm{100s}$ in fact increased between these two ages before decreasing relatively rapidly within a short period of time.
%-------------------------------------------
\subsection{Flaring luminosity $FA$}
\label{sec:fa}
We can relate the luminosity in flares in the Kepler band to the quiescent bolometric luminosity of the star, and by this define the fractional flare luminosity $FA$ (similar to \citetalias{ilin2019}):
\begin{align}
\label{eq:FA}
FA&=\dfrac{E_\mathrm{Kp,flare,tot,}}{t\cdot L_{\mathrm{bol,*,}}}
\end{align}
$E_\mathrm{Kp,flare,tot,}$ is the total energy released in flares, and $t$ is the total observing time of the star.
We determined $L_\mathrm{bol,*}$ from $R_*$ and $T_\mathrm{eff}$, as described in Section~\ref{TeffRL}. The energy released in flares was inferred using our derived stellar luminosities. By definition, $FA$ is a useful measure of relative stellar activity as long as the flux portion of the quiescent star in the Kepler band is roughly constant. One can compare $FA$ across age, but comparisons across $T_\mathrm{eff}$ bins should be interpreted with caution. $FA$ declines with age for every $T_\mathrm{eff}$ bin considered for both the total luminosity and relative to the quiescent flux~(Fig. \ref{fig:FA}). However, we found that $FA$ remained high ($>10^{-5}$) in early to mid M dwarfs for several hundred Myr after Pleiades age. 
\\ 
\citet{lurie2015} analysed the flares on GJ 1245 A and B, two M5 dwarfs in a triple system with another M8 dwarf, that were observed in 1 minute cadence during the original Kepler mission. GJ 1245 A and B both fall roughly into the $3000-3250\,$K bin, and rotate at periode $<1$\,d. The authors calculated $FA\approx 10^{-4}$ for both components, consistent with the $FA$ value in the 135 Myr old Pleiades in this $T_\mathrm{eff}$. But as the activity remains high at least up to the ages of Hyades and Praesepe, GJ 1245 could also be older.
\subsection{Hyades and Praesepe: Rotating differently}
\label{sec:hyaprarot}
%--------------------------------------------------------------------
   \begin{figure*}
   \centering
            \includegraphics[width=.85\hsize]{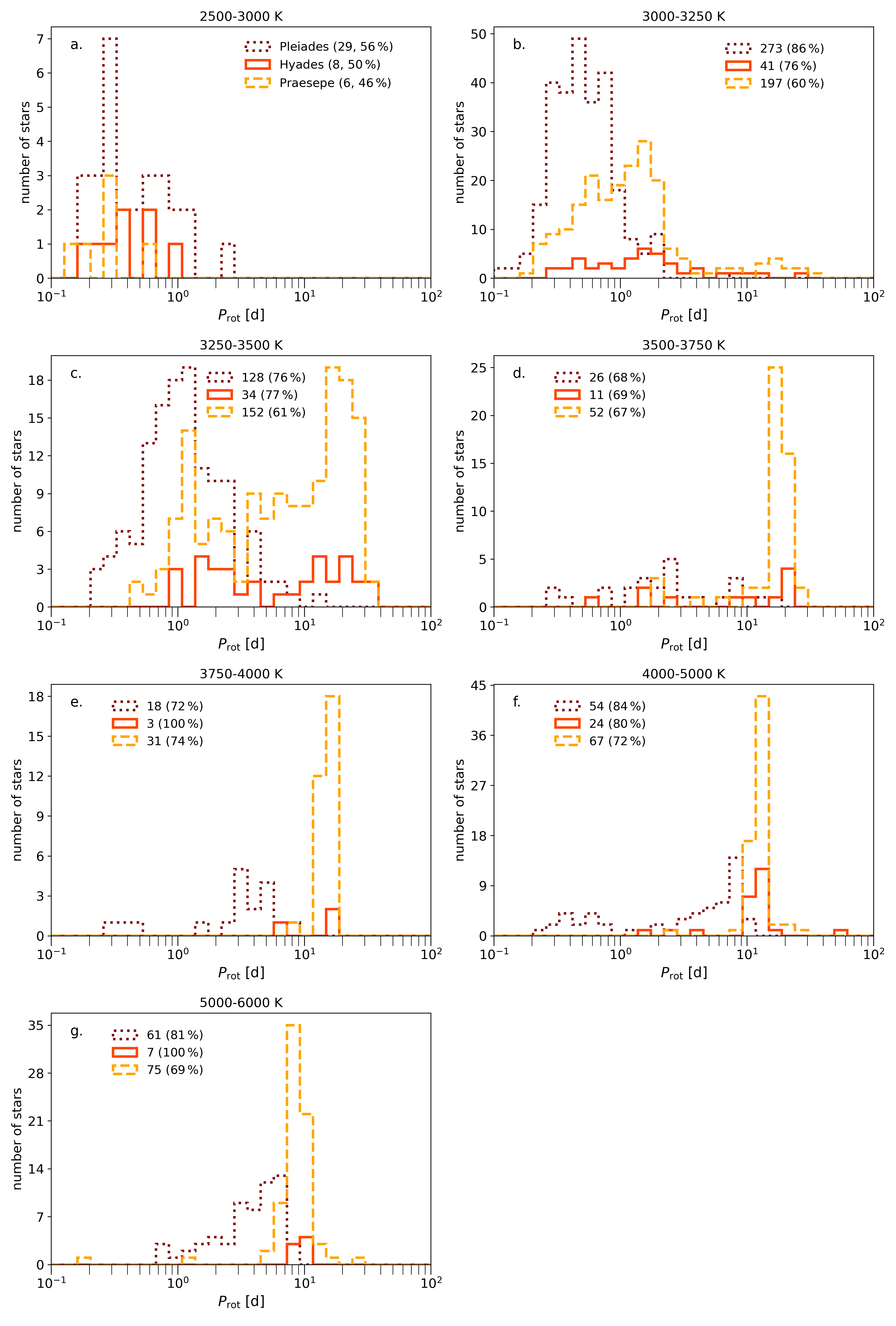}
         \caption{Rotation histograms for the cluster members that were searched for flares and had measured rotation periods $P_\mathrm{rot}$. Rotation periods were taken from~\citet{douglas2019} for the Hyades (orange solid) and Praesepe (yellow dashed), and from~\citet{rebull_pleiadesrot_2016} for the Pleiades (maroon dotted). The numbers and percentages in the legends in each panel denote the number of stars for which periods were given, and what fraction of the full sample that represented, respectively. The period distributions of Praesepe and the Hyades are consistent in all $T_\mathrm{eff}$ bins but the $3500-3750$ K bin, where the distribution of the Hyades appears more similar to the younger Pleiades than the same-age Praesepe. However, the sample size for the Hyades in that bin is small (11 stars).}
          \label{fig:rotation_histogram}
   \end{figure*}
%-------------------------------------------------------------------
We expected the flaring activity in the Hyades and Praesepe to be very similar because they are nearly coeval clusters with ages around 700 Myr, and comparable metallicities ([Fe/H](Praesepe)\,$=0.16$, [Fe/H](Hyades)\,$=0.13$,~\citealt{netopil_metallicities_2016}). We found this reflected in all our FFDs, except for the $3500-3750$\,K temperature bin, where the Hyades appeared almost as active as the Pleiades (135 Myr), while Praesepe showed flare frequencies about ten times lower than the Pleiades. This effect is not captured by the~\citet{davenport2019} model that predicts consistent flaring rates in all mass bins for Hyades and Praesepe~(Fig.~\ref{fig:powerlawfits_erg}). This discrepancy could be explained by the rotation period distribution in that bin. 
\\
In Fig. \ref{fig:rotation_histogram} we show rotation periods derived from K2 light curves for the Pleiades~\citep{rebull_pleiadesrot_2016}, the Hyades and Praesepe~\citep{douglas2019}, to illustrate the $P_\mathrm{rot}$ distributions that correspond to our $T_\mathrm{eff}$ bins.
\\
In all but the $3500-3750$\,K temperature bin the rotation period distributions of the Hyades and Praesepe were similar, while the Pleiades showed shorter periods on average. Only in this bin the rotation periods found in the Hyades and the Pleiades were more alike than in the Hyades and Praesepe. In this temperature regime, the majority of Praesepe stars had rotation periods $>10$\,d, while in the Hyades and the Pleiades rotation periods were more evenly distibuted in the $0.3-30$\,d range. A distinct transition in stellar activity was previously detected in various observables, including flaring luminosities and amplitudes, at the $P_\mathrm{rot}=10$\,d boundary~\citep{stelzer2016, lu2019}. Assuming that Praesepe is truly older than the Hyades~(see~\citet{douglas2019} for an indication of the opposite), the rotation period distribution appears to have made a jump between the ages of Hyades and Praesepe, and the flaring rates echoed this decrease. Considering the small sample size for the Hyades in this $T_\mathrm{eff}$ bin, a large spread in flaring activity levels is a second conceivable interpretation. In both cases, the results suggest a transition that occurred at the 10\,d boundary in the $3500-3750$\,K regime around the ages of the two clusters. A similar transition would then have already passed for hotter stars~(Fig.~\ref{fig:powerlawfits_erg} f.), and has yet to happen in lower mass stars~(Fig.~\ref{fig:powerlawfits_erg} b. and c.).
\subsection{A mass and rotation dependent flaring activity transition: Leaving the saturated regime}
\label{sec:4dim}
%-------------------------------------------
\begin{figure*}[ht!]
    \centering
    \includegraphics[width=\hsize]{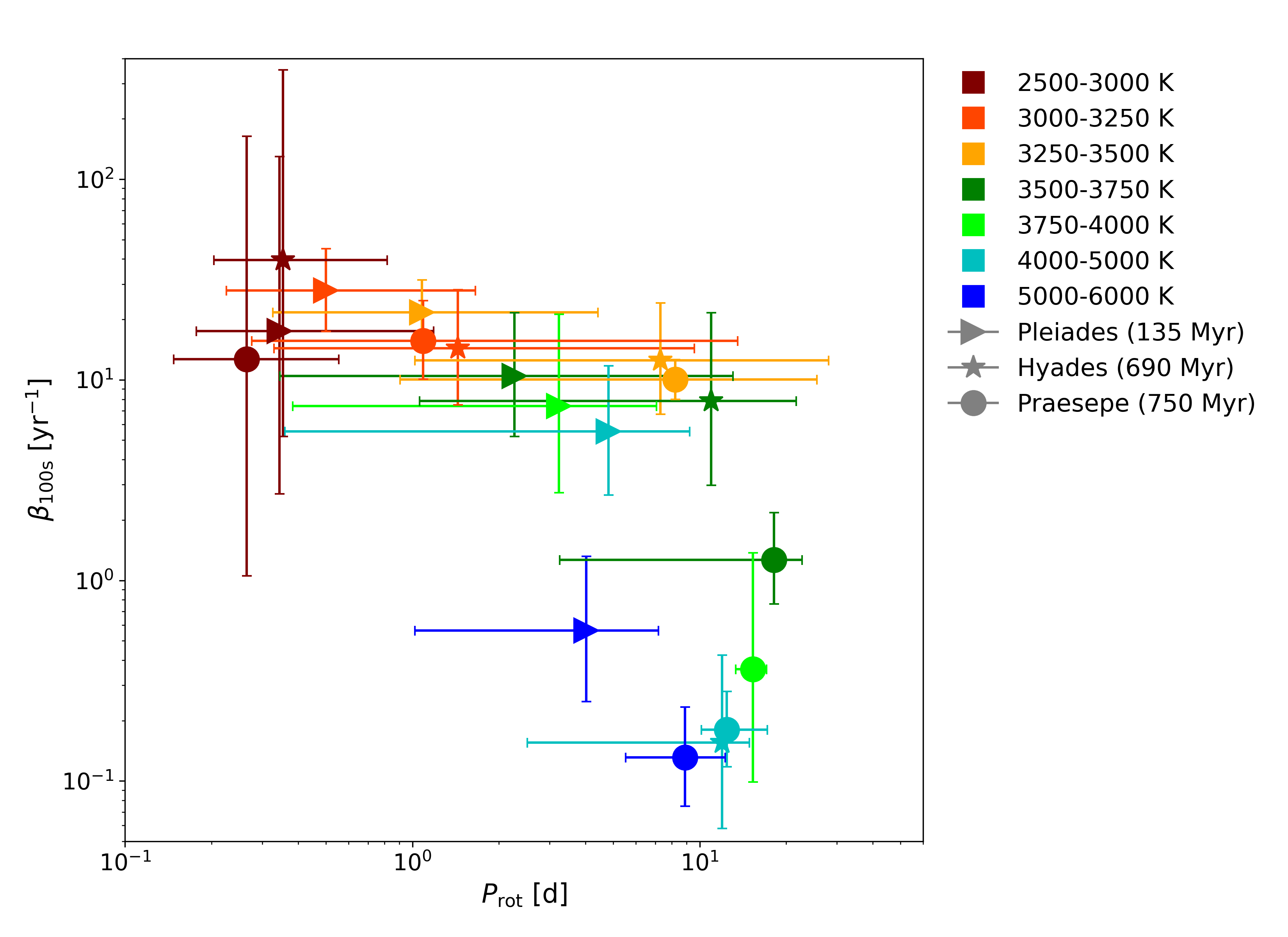}
    \caption{The power law fit intercept $\beta_\mathrm{100s}$ as a function of stellar rotation period $P_\mathrm{rot}$ (median value, and 5th to 95th percentiles of the distributions in Fig.~\ref{fig:rotation_histogram}). The saturated branch at high rotation rates and $\beta_\mathrm{100s}\gtrapprox 5$~yr$^{-1}$, and the drop in activity around the 10~d boundary resemble analogous distributions for X-ray, H$\alpha$, and surface magnetic field strength. The additional dimensions are age~(different symbols representing three open clusters), and $T_\mathrm{eff}$~(colors). At the ages of Hyades and Praesepe, the $T_\mathrm{eff}\sim 3500-3750$~K stars are appear to be in the process of leaving the saturated branch, while the $T_\mathrm{eff}\sim 3750-5000$~K stars already showed low activity levels. See Section~\ref{sec:4dim} for a detailed discussion.}       
    \label{fig:4dim}
\end{figure*}
%---------------------------------------------
To further explore the dependence of the flaring-age-mass relation on stellar rotation, we compared the distributions of rotation periods to $\beta_\mathrm{100s}$ in different age and $T_\mathrm{eff}$ bins~(Fig.~\ref{fig:4dim}). Up to Praesepe age, all stars below 3500 K showed a gradual decline in rotation speed with increasing $T_\mathrm{eff}$ while the flaring activity declined only marginally. These stars can be interpreted as residing in the saturated activity regime that is populated by young and rapidly rotating stars. This is consistent with age-rotation studies of surface magnetic field strengths~\citep{vidotto2014}, and a number of activity indicators, including coronal X-ray emission~\citep{pizzolato2003,wright2011}, and chromospheric H$\alpha$ and Ca II H and K lines~\citep{mamajek2008,west2015,newton2017}. In the $3500-3750$\,K temperature bin (see previous section), the median rotation rates changed as moderately as in the cooler bins. The Pleiades continued to spin down and gradually decreased in flaring activity. However, despite their similar age, the stars in the Praesepe sample spun down more than those in the Hyades, and were lower in $\beta_\mathrm{100s}$ by almost an order of magnitude. At higher temperatures up to $5000$\,K, the Pleiades remained on the saturated activity branch declining only gradually, while both Hyades and Praesepe reappeared at much lower activity levels, indicating a mass-dependent activity transition that had occurred in the meantime.
\\
In the highest mass bin ($5000-6000$\,K) the stars in Praesepe remained at similar rotation rates and activity levels as the $4000-5000$\,K stars. The highest mass Pleiades appeared to be an outlier with intermediate flaring rates at relatively short rotation periods. Solar-type stars at ZAMS are expected to show a wide spread in activity. In the simulations by~\citet{johnstone2020}, the slow and medium solar-type ZAMS rotators have left the saturated branch at 135 Myr, while the fast rotators remain highly active. As in the $3500-3750$\,K case, $\beta_\mathrm{100s}$ does not capture this intrinsic spread, but assumes similar activity levels for all stars in a given $T_\mathrm{eff}$-age bin. As a consequence, we measured intermediate flaring rates instead of spanning the range from high to low activities. We expect that this limitation applies to a much lesser degree to the other $T_\mathrm{eff}$ and ages, because of the general agreement between the nearly coeval Hyades and Praesepe, and because the stars in each of these bins are predicted to be found either completely on the saturated branch, or to have left it without exceptions~\citep{johnstone2020}.
%-------------------------------------------
\subsection{M67 and Ruprecht 147: Possibly binaries}
\label{sec:m67r147}
%-------------------------------------------
\begin{table}

\caption{Possible binary configurations for flaring old stars.}
\label{tab:m67_rup147_binaries}
\centering
\begin{tabular}{lccc}
\hline\hline
 cluster & EPIC & median SpT &     binary \\
\hline
     M67 &  211434440 &         K1 &    K2 + M5.5 \\
 Ruprecht 147 &  219601739 &         G8 &      K1 + M6 \\
 Ruprecht 147 &  219610232 &       K0.5 &    K2 + M5.5 \\
 Ruprecht 147 &  219591752 &         M3 &  M3.5 + M3.5 \\
\hline
\end{tabular}
\end{table}

%-------------------------------------------
\begin{figure}[ht!]
    \centering
    \includegraphics[width=\hsize]{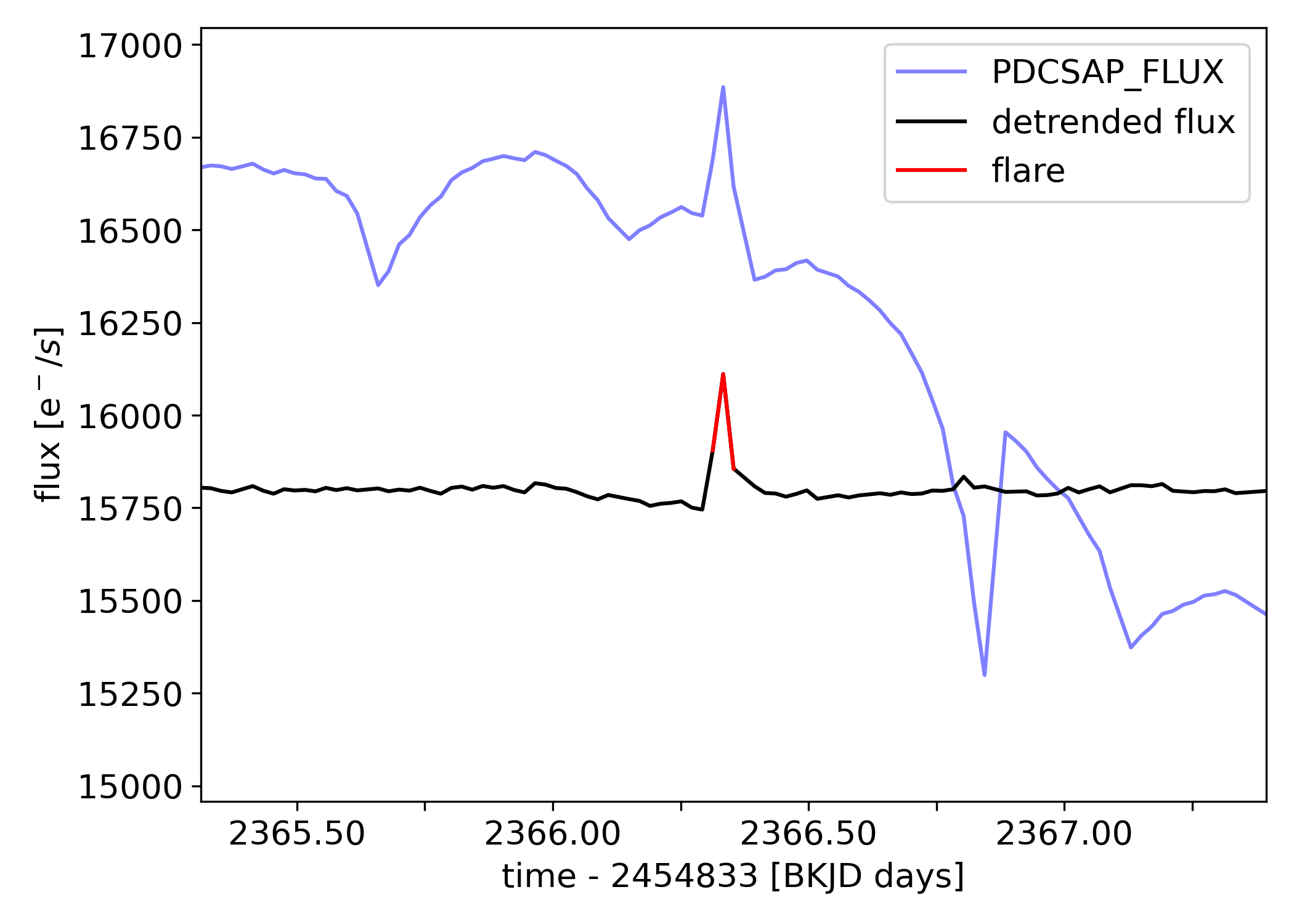}
    \caption{Single confirmed flare candidate in M67 on the K1 dwarf EPIC 211434440 (red line). Blue: Pre-search Data Conditioning (PDCSAP) flux, that is, the light curve before \texttt{K2SC} de-trending. Black: De-trended light curve that was searched for flares with \texttt{AltaiPony}. The amplitude of the flare is $2\%$, and the energy emitted in the Kepler band is $E_\mathrm{Kp}=6.25\cdot10^{33}$ erg. The flare is small enough to be barely detectable, so its energy is uncertain, and most likely underestimated due to the low time sampling of the light curve.}      
    \label{fig:m67}
\end{figure}
%-------------------------------------------
We found 8 and 4 flaring members in M67 and Ruprecht 147, respectively. Upon close inspection the majority were false positives, or occurred on multiple systems or evolved stars that were not properly filtered out. Most flares in these old clusters were detected on RS CVn binaries, cataclysmic binaries, spectroscopic binaries, and red giant stars. Excluding all these, we were left with one flare in M67 on an early K dwarf~(Fig. \ref{fig:m67}). In Ruprecht~147, we excluded EPIC 219426848, a double-lined spectroscopic binary~\citep{curtis2013}, and so narrowed down the list to a flare on a G8 star in Ruprecht~147, and four flares each on a K0.5 and an M3 star. For these stars, the multiplicity status was unknown. The uncertainty about the stellar properties suggests a range of possible masses of the system. The mass budget could be calculated from the uncertainties on their radii $R_*$ using the relations from \citet{eker2018}. In all cases, it was large enough that the stars in question could in principle be binary stars with mid-M dwarf companions that were too faint to be detected. We calculated hypothetic binary pairs for the cases where the primary mass was the smallest possible within $1\sigma$ on $R_*$. We give the median spectral type (SpT) of the target if it was a single star, and the spectral types of the possible binaries in Table~\ref{tab:m67_rup147_binaries}.
\\
After the manual inspection of the younger clusters we concluded that, while binaries and false positives were present in these clusters also, they were not the dominating source of flares. As multiplicity rates for low mass stars decrease from about 50\% for F type stars~\citep{raghavan2010} to 22\% for L and T dwarfs~\citep{duchene2013}, with M dwarfs showing somewhat higher multiplicity rates at about 27\%~\citep{winters2019}, we expect that the error introduced by unresolved binaries on our results was smaller for the cooler $T_\mathrm{eff}$ bins.
\section{Discussion}
\label{sec:discussion}
%-------------------------------------------
The launch of the Kepler satellite in 2009 caused a surge in statistical flare studies, a trend that TESS will continue and expand. Comparing our results with these works~\citep{shibayama2013,lurie2015,lin2019,raetz2020} and recent ground based surveys~\citep{chang2015,howard2019} we found encouraging consistency but also some noteworthy discrepancies that we could not always trace back to their causes~(Section~\ref{sec:consistency_other_work}). We discuss the disagreement between our FFDs and a gyrochronological model recently proposed for its parametrization as a function of mass and age~(Section~\ref{sec:davenport},~\citealt{davenport2019}). We conclude the discussion with a reflection on the observational limitations of the representation of FFDs by power law distributions~(Section~\ref{sec:sec:universal}), and the possibility of observing the maximum possible flare energy in our FFDs~(Section~\ref{sec:sec:maxen}).
\subsection{Consistency with statistical flares studies}
\label{sec:consistency_other_work}
%-------------------------------------------
\begin{figure*}[ht!]
    \centering
    \includegraphics[width=13.5cm]{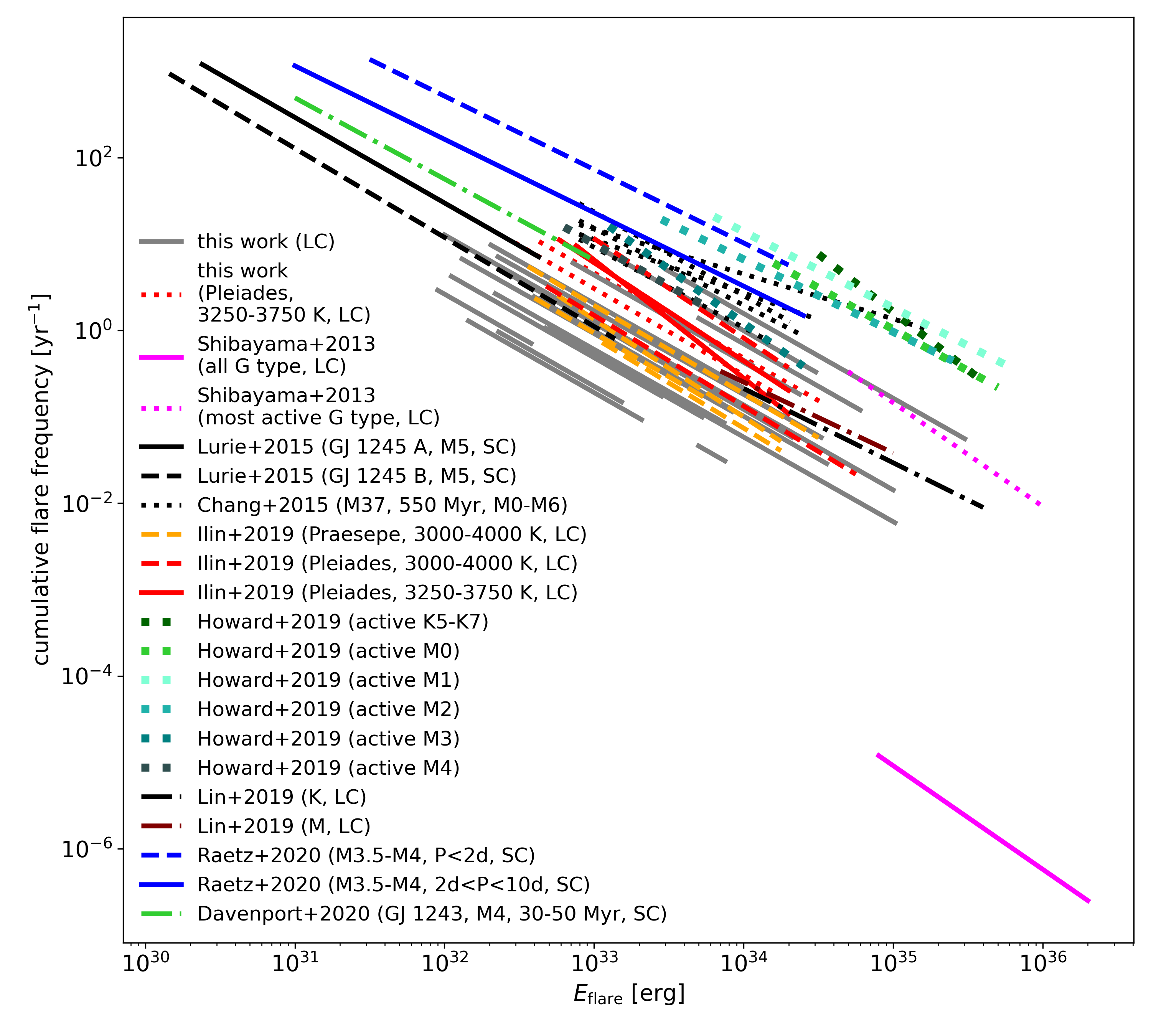}
    \caption{Comparison of flare frequency distributions (FFD) found in different flare studies. Gray lines: FFDs from this work. The one-flare sample for 5000-6000 K in M67 is indicated as the shortest visible gray line. Red solid lines, and red and orange dashes: K2 long cadence light curves study of the Pleiades and Praesepe in four $T_\mathrm{eff}$ bins from 3000 to 4000 K each~\citepalias{ilin2019}. Magenta line and dots: Superflares on all G dwarfs in Kepler, and its most active sub-sample~\citep{shibayama2013}. Black line and dashes: Two M5 stars in an M5-M5-M8 triple system observed by Kepler~\citep{lurie2015}. Black dots: Flare study based on an MMT survey of the $\sim550$ Myr old open cluster M37~\citep{chang2015}. Dots in green and blue: Evryscope all-sky flare search on late K to mid M dwarfs~\citep{howard2019}. Black and brown dash-dotted lines: K2 long-cadence study of K and M dwarfs within 200pc and 100pc~\citep{lin2019}. Dark blue line and dashes: K2 short cadence light curves of M3.5-M4 dwarfs with known rotation periods below $2\,$d and between $2$ and $10\,$d respectively~\citep{raetz2020}. Green dash-dotted line: The $30-50$\,Myr old M4 flare star GJ 1243 exhibited a constant flaring activity for over a decade of intermittent space based observations with Kepler and TESS~\citep{davenport2020arxiv}. In the studies that are based on Kepler or K2 observations, "SC" and "LC" denote the use of 1 minute and 30 minute cadence time series, respectively.}      
    \label{fig:otherwork}
\end{figure*}
%---------------------------------------------

%-------------------------------------------
\begin{figure}[ht!]
    \centering
    \includegraphics[width=\hsize]{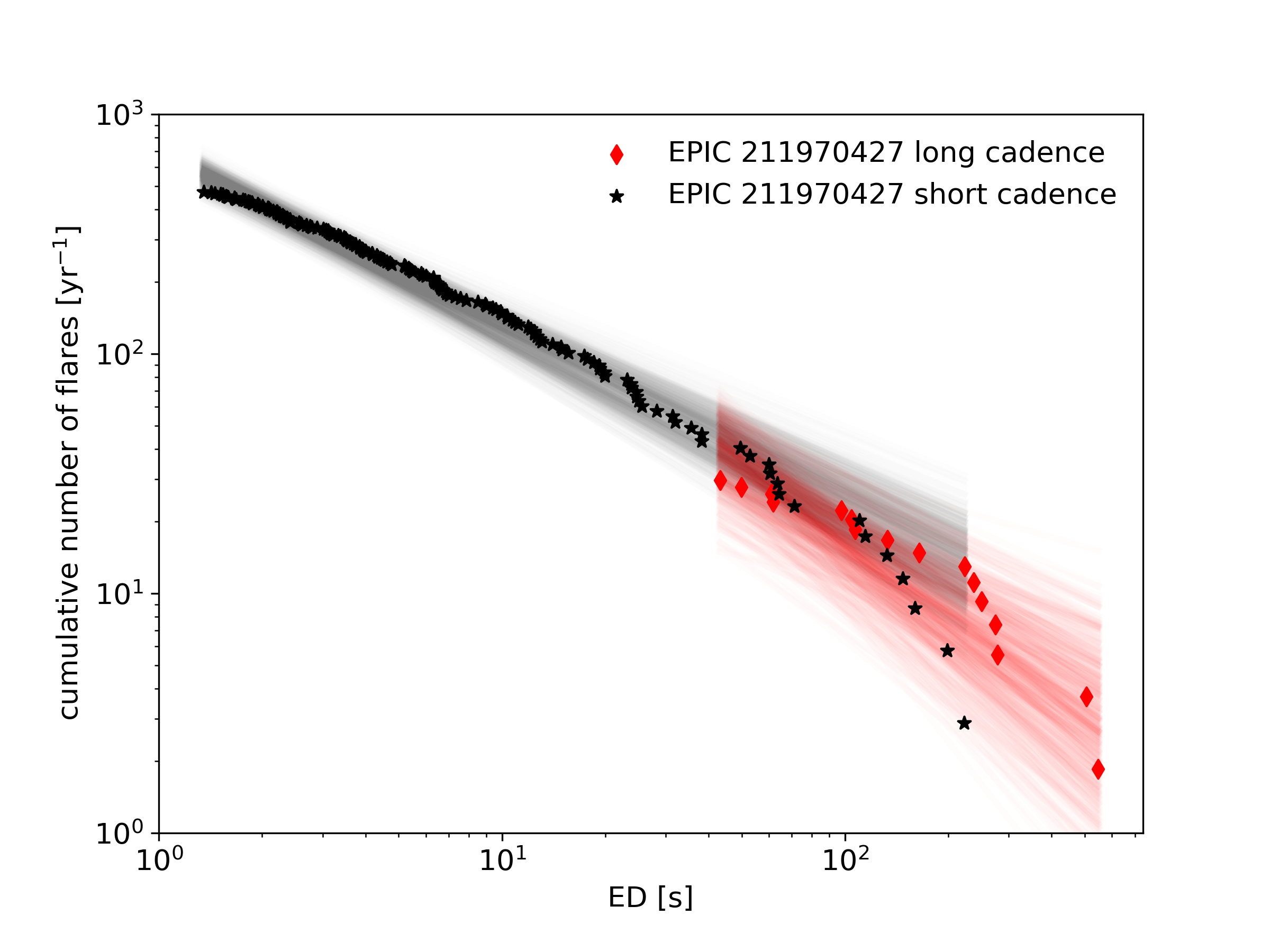}
    \caption{FFDs in $ED$ for EPIC~211970427, an M3 dwarf in Praesepe that was observed in both short 1 minute cadence (campaigns 5, 18) and long 30 minute cadence (campaigns 5, 16, and 18). We detrended, and searched both light curve types for flares using the method described in Section \ref{sec:sec:flarefinding}. Red diamonds indicate flares found in the long cadence light curves, black stars are flares from the short cadence light curves. The red and black lines were randomly sampled from the posterior distributions of the power law fits to the long and short candence FFDs, respectively~(see methods in Section~\ref{powerlawfits}).}          
    \label{fig:shortlong}
\end{figure}
%---------------------------------------------
We found our results to be broadly consistent with previous work both on Kepler/K2 data, and ground based surveys in the optical regime, with some noteable exceptions. Fig.~\ref{fig:otherwork} presents the FFDs of the studies that we discuss in this context. Overall, we believe that the differences between individual results were rooted in a combination of astrophysical properties like age, mass, rotation speed, and the influence of selection criteria on the respective samples. 
\\
The flare frequency distributions of two fast rotating (0.26 and 0.71 days) M5 flare stars in a M5-M5-M8 triple system~\citep{lurie2015} were consistent with the FFDs of flares in the Pleiades, Praesepe, and Hyades in the $3000-3250\,$K bin. 
\\
We also found that the FFDs of nearby K and M dwarfs, observed by Kepler/K2 within 200pc and 100pc, respectively, fell in the range between the Pleiades and Hyades/Praesepe FFDs in our sample~\citep{lin2019}.
\\
The frequency of superflares at $10^{34}$ erg in the most active G-dwarfs and Sun-like stars in Kepler was once in 10–100 days, and every 440 years on average solar-type stars~\citep{shibayama2013}. Benchmarking on the flare detected on a K1 dwarf in M67 in the G to early K bin (5000-6000 K), and accounting for the fraction of the total flare energy that the Kepler passband covers at a flare temperature of 10 000 K, we found the flaring rate at this energy to amount to about once every 100 years. Doing the same for the Pleiades stars yielded that ZAMS G to K dwarfs flare once per year at $10^{34}$ erg. Our young G-K stars therefore were less active than the active G dwarfs in \citet{shibayama2013}, while our solar-age G-K dwarfs appeared more active than in their results.
\\
\citet{howard2019} monitored superflares on cool stars with bolometric energies above $10^{33}$ erg and up to $10^{36}$erg. They found power law exponents $\sim 2$ in FFDs resolved by spectral type. The activity levels for spectral types K5 to M4 were on average higher than in our sample. However, \citet{howard2019} only included stars that exhibited flares (active stars) into their FFDs, whereas we also accounted for the observing time of stars where no flares had been detected.
\\
Photometric flares observed by the MMT 6.5 telescope~\citep{hartman2008} in the $\sim550$ Myr old open cluster M37~\citep{chang2015} appeared on average more active than the Pleiades in our study for spectral types M0-M6. This could imply that flaring activity in these low mass stars peaked not at Pleiades age but at a later evolutionary stage. 
\\
Similarly, \citet{raetz2020} pointed out that their fast rotating M1.5-M4.5 dwarfs appeared more active than the results for the Pleiades in \citetalias{ilin2019} in the $3250-3750\,$K range. Our revised results remained consistent with \citetalias{ilin2019}~(Fig. \ref{fig:otherwork}). Both samples were comprised of flares with periods below 10\,d. Rotation periods $P_\mathrm{rot}$ were given for 100\% of the sample in~\cite{raetz2020}, and for \mbox{$>68\%$} in our Pleiades sub-sample~(Fig. \ref{fig:rotation_histogram}). 
\\
We consider three plausible reasons for the discrepancy here. First, the M1.5-M4.5 dwarfs in \citet{raetz2020} could be at an age at which stars show higher flaring activity than stars with the same rotation periods in the Pleiades. Second, there may be selection effects in the short cadence sample in \citet{raetz2020} as compared to our sample. Targets for short cadence observations in K2 were not selected randomly from the underlying stellar population, but were filtered for a variety of properties that biased the final selection toward more active stars. Third, our $3250-3750\,$K bin corresponds to spectral types M1-M3.5 rather than to M1.5-M4.5~\citep{pecaut_intrinsic_2013}. So the $T_\mathrm{eff}$ bins chosen in~\citet{raetz2020} encompassed later, and therefore more active, spectral types. A resolution of this discrepancy must also take into account that the results for our Praesepe sample in the  $3250-3750\,$K range were consistent with the slow rotators ($P_\mathrm{rot}\sim32\,$d) in \citet{raetz2020}.
\\
Finally, we note that the difference in time sampling between our study (30 minute candence), and 1 minute cadence in~\citet{raetz2020} can most likely be ruled out as an explanation for the discrepancy. In Fig.~\ref{fig:shortlong}, we show an example that illustrates that short cadence light curves used in~\citet{raetz2020}, and long cadence light curves studied here yielded FFDs that were consistent withing uncertainties. There were no exactly matching events in the FFDs although a subset of the short cadence flares was detected in long cadence also for two main reasons. First, in long cadence, flare energy was typically underestimated~\citep{yang_flaresampling_2018}. Second, smaller flares were merged together if they occurred within a sufficently short time period. What looked like one flare in long cadence was sometimes resolved into two or more flares in short cadence.
\subsection{Gyrochronological parametrization of flare frequency distributions}
\label{sec:davenport}
\citet{davenport2019} derived a parametrization of the FFD arguments $\alpha$ and $\beta$ as a function of mass and age using the Kepler flare catalog~\citep{davenport_kepler_2016}, rotation periods from~\citet{mcquillan2014}, and the \citet{mamajek2008} gyrochrone model. We show the model power laws in Fig. \ref{fig:powerlawfits_erg} alongside our results. While the power law exponents appeared consistent with our estimates, the activity level in our sample was much lower than predicted by the model. A deviation was expected as the model was not designed to predict absolute ages. \citet{davenport2019} noted a sample bias in their study toward more active stars. As discussed by the authors, their model overpredicted the superflaring rate of the average Sun-like sample from~\citet{shibayama2013} and more resembled the rate for their most active sub-sample. Moreover, the model was not designed to be applicable to flaring activity in the saturated regime of fast young rotators~\citep{jackman2020}. With our data we quantify the overestimation of flaring rates to be one to two orders of magnitude. Our results also suggest that the model overpredicts the activity level for old stars more than for younger stars~(Fig. \ref{fig:powerlawfits_erg}). We propose that future models include the mass-dependent drop in activity at $P_\mathrm{rot}\approx 10\,$d~(see Section~\ref{sec:4dim}), for instance, using a piecewise definition of the setup in~\citet{davenport2019} for stars with rotation rates above and below the 10\,d threshold.
\subsection{Universality of the power law exponent $\alpha$}
\label{sec:sec:universal}
%--------------------------------------------------------------------
   \begin{figure}[t!]
   \centering
            \includegraphics[width=\hsize]{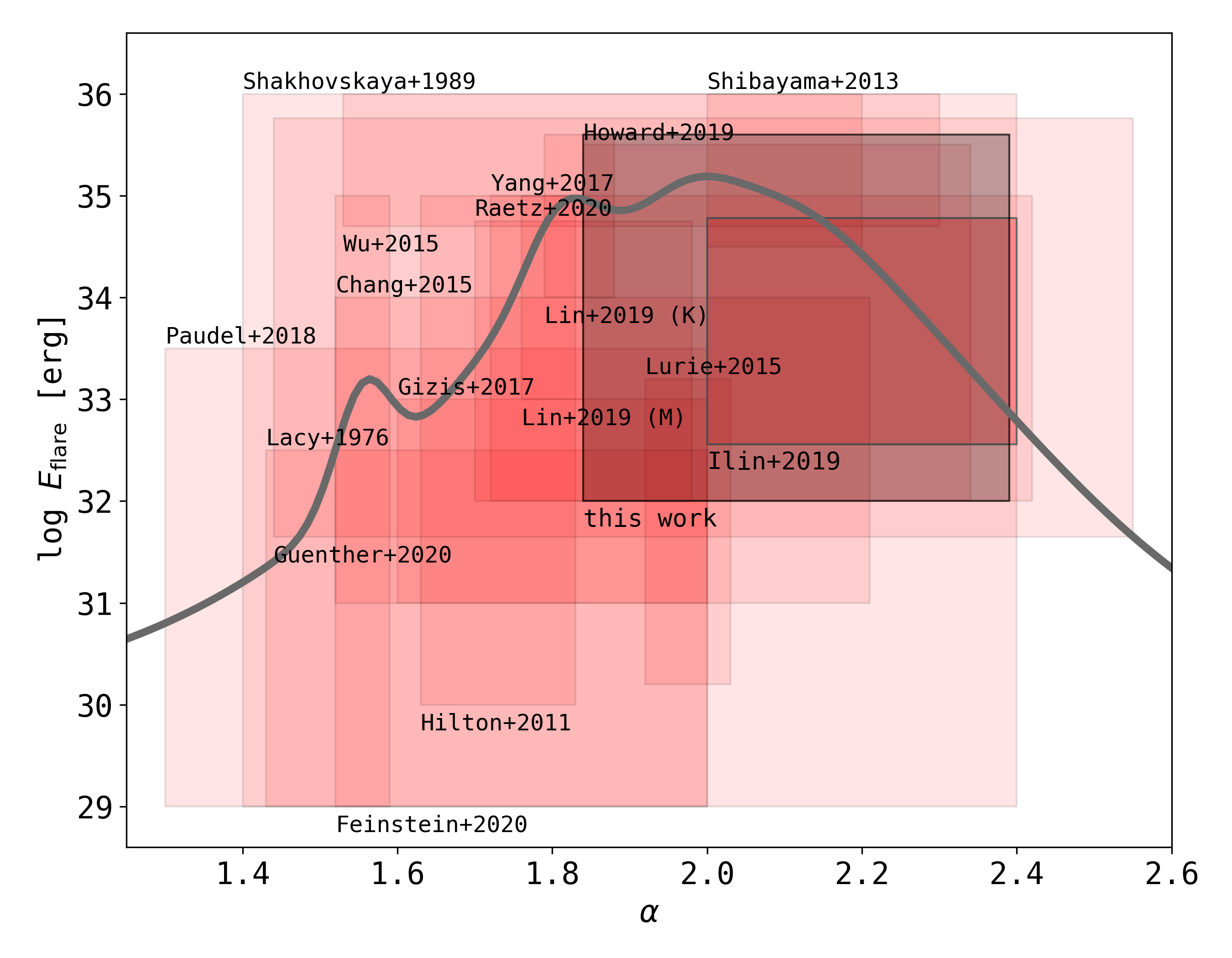}
         \caption{Literature overview over power law fits to FFDs. Red rectangles indicate the energy range in which flares were detected (height), and the range of values determined for the power law exponent $\alpha$~(width) including quoted uncertainties and superimposing the ranges for $\alpha$ from multiple FFDs if any. The references are anchored to the upper or lower left corner of the respective rectangle. The gray rectangle refers to this study. The gray line depicts the superposition of these results, representing each by a Gaussian fit with the range used as standard deviation, and weighted by the square root of the number of flares found in each study. \citetalias{ilin2019} is shown with a dark outline. See Section~\ref{sec:sec:universal} for details on the quoted works.
}
          \label{fig:powerlaw_literature}
     
   \end{figure}
We found values for $\alpha$ to range from 1.84 to 2.39~(Table~\ref{tab:powerlawtable_spt}), with smaller numbers of flares in a FFD usually leading to values closer to the margins of this range, and FFDs that consisted of more than 50 flares being better constrained with $\alpha\sim 1.89-2.11$.
Taking into account uncertainties and systematic errors resulting from the use of different power law fitting methods~\citep{maschberger2009}, the power law exponent $\alpha\sim 2$ appears to be similar for the majority of studies on flare statistics in the literature so far, irrespective of spectral type or age. To illustrate this, we searched the literature for studies where power laws were fitted to FFDs of flares in the optical regime using different methods~(Fig.~\ref{fig:powerlaw_literature}). 
\\
 \citet{lacy_uv_1976} pioneered in statistical flare studies using observations of eight UV Ceti stars. \citet{shakhovskaya_stellar_1989} analyzed monitoring data from several dozen active early K to late M dwarfs in the solar vicinity, the Orion, the Pleiades, the Hyades, and the Praesepe cluster. A number of studies focused on solar-type and G dwarfs stars in Kepler~\citep{shibayama2013,wu2015}. \citet{hilton2011} derived $\alpha$ from ground-based observations of M dwarfs. \citet{lurie2015} determined $\alpha$ from Kepler short cadence light curves of two active M5 dwarfs in Kepler. \citet{yang_flaring_2017} followed up on \citet{davenport_kepler_2016} in searching the entire Kepler catalog for flares. \citet{lin2019} (M) and (K) were based on nearby M and K dwarf flares in K2 long cadence light curves. \citet{raetz2020} studied short cadence M dwarfs with rotation periods in K2. \citet{gizis_k2_2017-1} constructed an FFD for 22 flares on an M7 dwarf. \citet{paudel2018} studied M6-L0 dwarfs in K2 short cadence data. \citet{howard2019} found over 500 superflares on 284 stars in the framework of the Evryscope all-sky survey. Photometric flares were observed by the MMT 6.5 telescope~\citep{hartman2008} in the $\sim550$ Myr old M37 open cluster~\citep{chang2015}. \citet{guenther2020} took a first look to $\alpha$ as derived from stars observed during TESS Cycle 1. From their work we only use $\alpha$ derived from stars with 20 or more flare detections in Fig.~\ref{fig:powerlaw_literature}. Finally, \citet{feinstein2020} used a convolutional neural network setup to detect flares in nearby young moving groups, open clusters, OB associations, and star forming regions with ages between 1 and 800 Myr in TESS, finding relatively flat FFDs compared to~\citet{guenther2020}.
       \\
While most studies consistently found values somewhat above or below $\alpha\approx2$, the comparison would reveal unresolved systematic errors in all these studies, including our own, if $\alpha$ was truly universal for all FFDs. The determination of $\alpha$ using different methods without reliable uncertainty estimates across different studies makes it difficult to assess whether the spread in $\alpha$ from about $1.4$ to $2.5$ is physically motivated or not. 
\\
Addionally, for both low and high flare energies, the detection of flares is usually incomplete or contaminated. This is caused by finite time sampling, cosmic ray impact, and scatter in the light curve at the low end, and finite observation time at the high end~(see Section \ref{sec:sec:injrec}). As long as these methodological and observational caveats apply, to answer the question about the universal nature of $\alpha$ proves very challenging.
\subsection{Maximum flare energy}
\label{sec:sec:maxen}
Spots can survive on the stellar surface from a few days to nearly a year~
\citep{namekata_solarstellarwlf_2017, davenport_flaresandspots_2015}. Complex spot geometry is correlated with the stongest X-class flares on the Sun~\citep{toriumi_flaresspotssun_2017, sammis_deltaspotsflares_2000}. These observations support the idea that flares are associated with the presence of certain types of starspots, or more generally, certain types of active regions. Since we expect that there is a maximum flare energy a spot can produce, the underlying power law relation must break above some maximum value in any FFD. We tested a possible truncation of our FFDs using the Kolmogorov-Smirnov test for consistency with a power law distribution, followed by the exceedance statistic to test for a truncation of this power law as suggested by~\citet{maschberger2009}. While most FFDs with $\geqslant50$ flares were consistent with a power law distribution, we found no evidence for truncation in any of the $T_\mathrm{eff}$-age bins. This would imply that we did not sample the highest possible energies. Such a conclusion is plausible because flare energies higher than our observed maximum flare energies were detected on other stars~(see Fig.~\ref{fig:otherwork},~\citealt{howard2019, lin2019, shibayama2013}). In a few FFDs, we still noted a deviation from a single power law at the high energy end. This effect could be caused by the combination of stars with different flaring rates into a single distribution, which would violate the assumption of similarity between the stars in each bin~(see Section~\ref{powerlawfits}).
\section{Summary and conclusions}
\label{sec:summary}
In this study, we investigated how flaring activity unfolded for different age and $T_\mathrm{eff}$ ranges anchoring the age to the membership in five open clusters. We found flaring activity to decrease from mid M dwarfs to G stars, and from ZAMS to solar age, except for stars later than M5, where activity increased from ZAMS to 690\,Myr. We found indications of a mass and rotation dependent threshold age, above which stars depart from the saturated flaring activity branch.
\\
Using multiple cluster membership studies we selected for high-probability members, and drew from a host of multiband photometry catalogs and Gaia astrometry to determine $T_\mathrm{eff}$ and stellar luminosities. Keeping only targets with well-determined mass and age we proceeded to search 3435K2 long cadence light curves obtained for these stars throughout the K2 mission. We applied the open source software \texttt{K2SC} to remove rotational variability and instrumental effects. We developed the open source software \texttt{AltaiPony}, and used it to automatically detect and determine the properties of flare candidates. We vetted all flare candidates manually and discussed various sources of false positives and incompleteness on both the high and the low energy end of the FFDs. All in all, we searched 2111stars in five open clusters, and found 3844flares. Most flares originated in the Pleiades and Praesepe, several hundred were found in the Hyades, a handful in Ruprecht 147, and only one flare candidate appeared on a K1 dwarf in M67, although several flares were found in binary systems in both Ruprecht 147, and M67.
\\
We fitted power law functions to the flare frequency distributions of flares from stars in different age and $T_\mathrm{eff}$ bins. We found that the power law exponent $\alpha$ was consistent with previous work, and caution against interpreting apparent trends, especially when comparing across different studies.
\\
Our results indicated that flaring activity declined with age, and did so faster for higher mass stars. We provide the measured flaring rates in Table~\ref{tab:powerlawtable_spt} as a contribution to the endeavor of mapping the flaring-age-mass relation from pre-main sequence to main sequence turn-off, and from the onset of an outer convection zone in F-type stars to the coolest brown dwarfs.
\\
Our findings paint a picture of clear trends in the flaring evolution of GKM dwarfs on the main sequence. Regardless of varying methods employed for flare search and FFD analysis in previous studies we found encouraging consistency in flaring rates with both Kepler and K2 based work, and ground-based surveys. We noticed some differences, which we could not always resolve as being either systematic or astrophysical. We suggest that discrepancies between our results and flare studies that used rotation periods for their age estimates~\citep{davenport2019, raetz2020} could be explained by sample selection bias but may also point to limitations of rotation periods as an age indicator. Noteably, we found the flaring activity in M1-M2 dwarfs in Hyades and Praesepe to differ significantly despite their similar ages, while it was consistent for the two clusters for both higher and lower mass stars. We interpret this as a sign of a mass and rotation dependent transition from saturated to unsaturated activity levels: The flaring rates of higher mass stars at somewhat lower rotation periods had already declined substantially, while lower mass stars still resided in the saturated activity regime, and were spinning faster. We conclude that some discrepancies between our results and flare studies that used rotation periods for their age estimates could be explained by sample selection bias toward more active stars, but others may hint at limitations to using rotation as an age indicator without additional constraints from stellar activity. 
\\
The impact of flares from low-mass stars on rocky exoplanets in the habitable zone is a key component to understanding the space weather environments in which life may develop on these objects. Flares are promising candidates for being a major source of high energy radiation~\citep{airapetian2020} that can fertilize the emergence of biological life or impede it if overdosed; it may sweep away or alter the composition of these planets' atmospheres, and even evaporate oceans full of surface water~\citep{shields2016, tilley2019}. Age-calibrated flare frequency distributions that trace the entire lifespan of planet hosting stars are indispensable ingredients to the assessment of planetary atmosphere survival under repeated flaring. 
\\
Although Kepler and K2 data were already used in more than 2\,400 publications in 2018, the public archive was still considered understudied~\citep{barentsen_retirement_opportunities_2018}. It will remain a benchmark mission not only for exoplanet search, but also for various subfields in stellar astronomy, and flaring activity research, in particular. Results from the first year of operations of TESS indicate that the light curve quality delivered by the mission so far was well suited for statistical flare studies, too~\citep{doyle2020,guenther2020, feinstein2020}. The expected lifetime of TESS could be up to 20 years, in which case the number of high quality flare samples will soon outclass Kepler and K2 as the primary source of flares. The mission will vastly expand the treasury of light curves, leaving us optimistic that this work was only a first step toward a comprehensive flare rate evolution model. 
\begin{acknowledgements}
EI is thankful to the anonymous referee for their detailed, constructive, and respectful report, Michael Gully-Santiago for invaluable advice in the development of \texttt{AltaiPony}, Ann-Marie Cody for helpful comments on K2 data, Nikoleta Ilić for assisting with code review, and Klaus Strassmeier for helpful remarks on the manuscript. EI acknowledges support from the German National Scholarship Foundation. JRAD acknowledges support from the DIRAC Institute in the Department of Astronomy at the University of Washington. The DIRAC Institute is supported through generous gifts from the Charles and Lisa Simonyi Fund for Arts and Sciences, and the Washington Research Foundation.
This paper includes data collected by the Kepler mission and obtained from the MAST data archive at the Space Telescope Science Institute (STScI). Funding for the Kepler mission is provided by the NASA Science Mission Directorate. STScI is operated by the Association of Universities for Research in Astronomy, Inc., under NASA contract NAS 5–26555.
We made use of data products from the Two Micron All Sky Survey, which is a joint project of the University of Massachusetts and the Infrared Processing and Analysis Center/California Institute of Technology, funded by the National Aeronautics and Space Administration and the National Science Foundation. The Pan-STARRS1 Surveys (PS1) and the PS1 public science archive have been made possible through contributions by the Institute for Astronomy of the University of Hawaii, the Pan-STARRS Project Office, the Max-Planck Society and its participating institutes, the Max Planck Institute for Astronomy, Heidelberg and the Max Planck Institute for Extraterrestrial Physics, Garching, The Johns Hopkins University, Durham University, the University of Edinburgh, the Queen’s University Belfast, the Harvard-Smithsonian Center for Astrophysics, the Las Cumbres Observatory Global Telescope Network Incorporated, the National Central University of Taiwan, the Space Telescope Science Institute, the National Aeronautics and Space Administration under Grant No. NNX08AR22G issued through the Planetary Science Division of the NASA Science Mission Directorate, the National Science Foundation Grant No. AST-1238877, the University of Maryland, Eotvos Lorand University (ELTE), the Los Alamos National Laboratory, and the Gordon and Betty Moore Foundation.
This research also made use of the Python packages \texttt{numpy}~\citep{numpy2011}, \texttt{pandas}~\citep{pandas2010}, \texttt{astroML}~\citep{astroML2012, astroML2014}, \texttt{astropy}~\citep{astropy2013}, \texttt{specmatch-emp}~\citep{yee_specmatch_2017}, \texttt{bokeh}~\citep{bokeh}, the cross-match service provided by CDS, Strasbourg; IMCCE's SkyBoT VO tool; data from the European Space Agency (ESA) mission {\it Gaia} (\url{https://www.cosmos.esa.int/gaia}), processed by the {\it Gaia} Data Processing and Analysis Consortium (DPAC, \url{https://www.cosmos.esa.int/web/gaia/dpac/consortium}). Funding for the DPAC has been provided by national institutions, in particular the institutions participating in the {\it Gaia} Multilateral Agreement.
\end{acknowledgements}
\bibliography{MyLibrary}
%\Online
\begin{appendix}
%--------------------------------------------------------------------
\section{Membership probabilities}
\label{app:memberships}
\begin{figure*}[ht!]
            \includegraphics[width=\hsize]{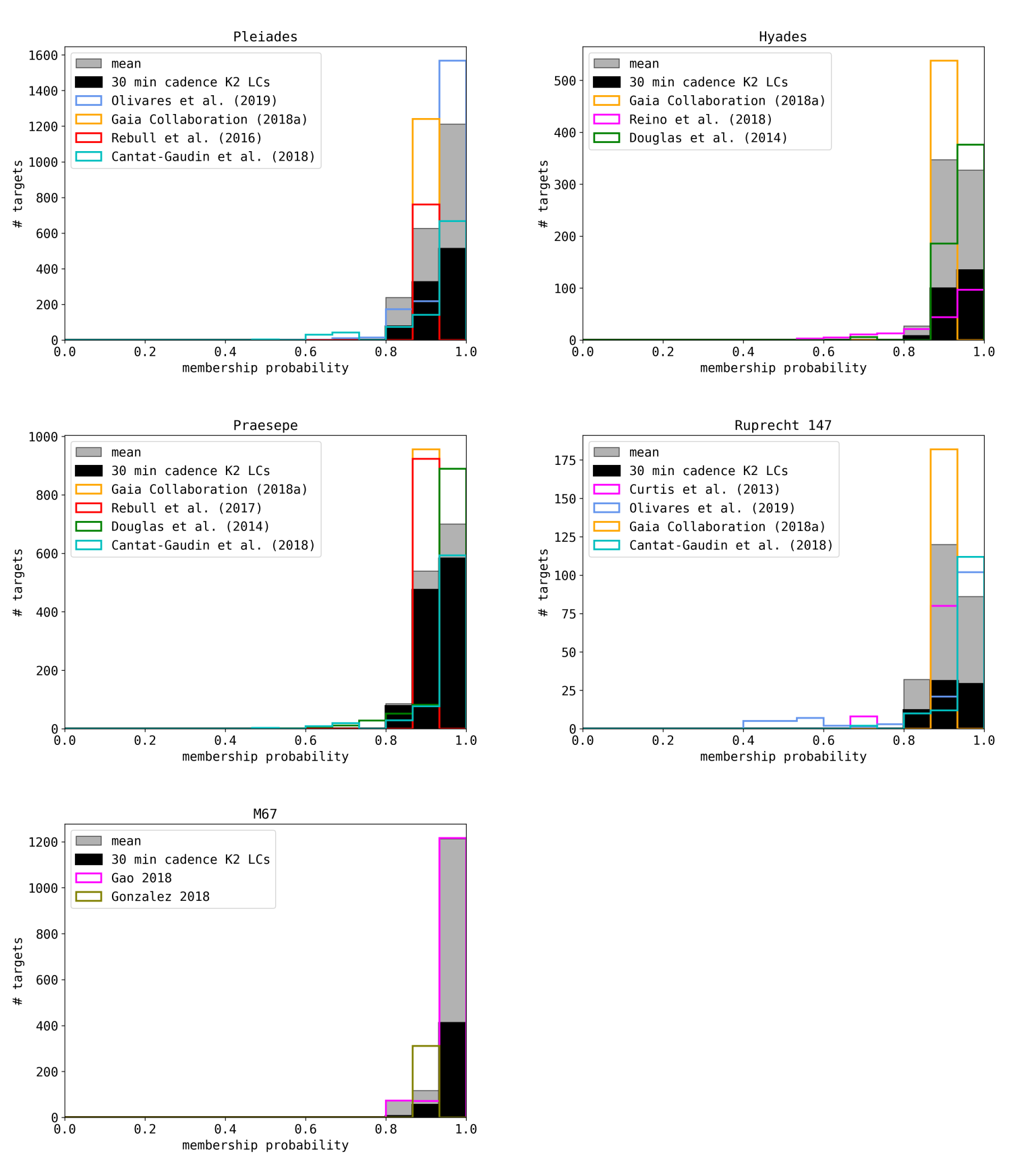}
         \caption{Open cluster membership histograms and selected targets. Each panel corresponds to the distribution of candidate members of the cluster indicated in the panel title. Membership catalogs are shown as step histograms (different colors). The mean membership probability distribution is a filled histogram (gray). For our study, we selected only stars with membership probabilities above 80\% that were observed by K2 in long cadence mode (black filled histogram).}
          \label{fig:app:memberships}

   \end{figure*}

To match catalogs on RA and declination we used the \texttt{astroML.crossmatch} tool for Python~\citep{astroML2012}.
\\
For the studies with classifiers we assigned membership probabilities as follows.
In~\citet{gonzalez_m67mem_2016}:
\begin{eqnarray*}
p(M (\text{member}))&=&0.9,\\
p(BM(\text{binary member}))&=&0.9,\\
p(N (\text{non-member}))&=&0.1,\\
p(SN(\text{single non-member}))&=&0.1,\\
p(BN (\text{binary non-member}))&=&0.1,\\
p(U (\text{unknown member}))&=&0.5.
\end{eqnarray*}
In~\citet{curtis2013}:
\begin{eqnarray*}
p(Y (\text{highest confidence member}))=0.9,\\
p(P (\text{possible/probable member}))=0.7,\\
p(N (\text{not likely/non-member}))=0.7,\\
p(B (\text{photometry consistent with blue stragglers}))=0.0.\\
\end{eqnarray*}
In~\citet{rebull_praesepe_2017}:
\begin{eqnarray*}
p((\text{best}))=0.9,\\
p((\text{ok}))=0.6,\\
p((\text{else}))=0.1.
\end{eqnarray*}
Members from \citet{rebull_pleiadesrot_2016, douglas_poking_2017}; and ~\citet{gaia_dr2_2018_hrd} were assigned $p=0.9$ if they appeared in the final catalog.
\\
Table \ref{tab:app:memberships} gives an overview over different membership catalogs. Figure \ref{fig:app:memberships} shows membership probability histograms of the final sample broken down by membership source.

%--------------------------------------------------------------------

\begin{table*}
\caption{Membership catalogs used in this work.}
\label{tab:app:memberships}
\centering
\begin{tabular}{llll}     % 7 columns
\hline\hline
     source  & type  & clusters covered & notes\\
\hline
   \citet{curtis2013} & classifier & Ruprecht 147 & \\
   \citet{douglas_praesepe_hyades_2014} & probability & Hyades, Praesepe & meta study \\
   \citet{gonzalez_m67mem_2016} & classifier & M67 & \\
   \citet{rebull_pleiadesrot_2016} & members list & Pleiades & meta study\\
   \citet{rebull_praesepe_2017} & classifier & Praesepe & meta study\\
   \citet{douglas_poking_2017} & members list & Praesepe & meta study\\
   \citet{gaia_dr2_2018_hrd} & members list & Hyades$^{\dagger}$,   & Gaia DR2\\
   &&Ruprecht 147, Pleiades, &\\
   &&Praesepe&\\
   \citet{cantat_gaudin_2018} & probability & Ruprecht 147, & Gaia DR2\\
   && Pleiades, Praesepe&\\
   \citet{gao_m67mem_2018} & probability & M67 & Gaia DR2\\
   \citet{reino_hyades_2018} & probability & Hyades$^{\dagger}$ & Gaia DR1\\
   \citet{olivares_pleiades_2018} & probability & Pleiades & Gaia DR2, DANCe\\
   \citet{olivares_ngc6774_2019} & probability & Ruprecht 147 & Gaia DR2, DANCe\\
\hline
\end{tabular}
\tablefoot{Meta study: See references in the studies for additional membership information. DANCe: DANCe membership study project.  Gaia DR1: The study used data from the first data release~\citep{gaia2016_2} of the Gaia mission~\citep{gaia2016}. Gaia DR2: The study used data from the second data release~\citep{gaia2018}. $^{\dagger}$Positions for Hyades were propagated to epoch 2000 using Gaia proper motions.}
\end{table*}

%--------------------------------------------------------------------

%--------------------------------------------------------------------
\section{Cluster parameters}
\label{app:cluster_params}
We compiled various age, distance, and metallicity values from the literature for the clusters in this study. Table \ref{tab:app:oc_parameters} shows that most studies independently arrived at similar results for all our clusters. The adopted parameters are emphasized, and were chosen based on consistency with prior work, uncertainties provided with the results and the data used for the study (newer catalogs were preferred). 
\begin{table*}
\caption{Non-exhaustive literature overview over OC parameters.}
\label{tab:app:oc_parameters}
\centering
\begin{tabular}{lllll}
\hline\hline
     cluster &                                source & distance [pc] &                            age [Myr] &                       [Fe/H] \\
\hline
\textbf{Pleiades} & \textbf{adopted in this work:}   &     135.6     &  $ 135       \pm _{ 25}^{25       }$ & $ -0.037        \pm 0.026 $  \\
	    &\citet{bossini2019} \tablefootmark{a}  && $86.5 \pm _{ 2.4}^{6       }$        &                              \\
 	    &           \citet{cantat_gaudin_2018} &     135.6     &                                      &                              \\
 	    &             \citet{gossage2018}     &               &  $ 135       \pm _{ 25}^{25       }$ &                              \\
	    &            \citet{yen2018}            &     126.3     &  $ 141.3     \pm _{ 100}^{170     }$ &                              \\
 	    &            \citet{chelli2016}   &     139       &                                      &                              \\
 	    &  \citet{netopil_metallicities_2016}   &               &                                      &               -0.01          \\
	    &             \citet{dahm_reexamining_2015}              &               &  $ 112       \pm _{ 5}^{5         }$ &                              \\
 	    &            \citet{scholz2015}     &     130       &                           120        &                              \\
	    &        \citet{conrad2014}    &               &                                      &  $ -0.037        \pm 0.026 $ \\
	    &             \citet{melis2014}      &     136       &                                      &                              \\
			  &   \citet{bell_pre-main-sequence_2012}   &     135       &                           125        &                              \\\hline
\textbf{Hyades} & \textbf{adopted in this work:\tablefootmark{c}}     &               &  $ 690       \pm _{ 100}^{160     }$ &  $0.13  \pm 0.02$            \\	       	&             \citet{gaia_dr2_2018_hrd} &               &  $ 690       \pm _{ 100}^{160     }$ &                              \\
       &             \citet{gossage2018}     &               &                           680        &                              \\
       &             \citet{liu2016}        &               &                                      &  $               \pm 0.02  $ \\
       &             \citet{netopil_metallicities_2016}    &               &                                      &               0.13           \\
       &             \citet{taylor2005}  &               &                                      &  $ 0.103         \pm 0.008 $ \\
 %Hyades      &             Cummings et al. (2005)    &               &                                      &  $ 0.146         \pm 0.004 $ \\
       &            \citet{salaris_age_2004}    &               &  $650$                               &  $ 0.15$                      \\
       &             \citet{perryman1998}    &               &  $ 625       \pm _{ 50}^{50       }$ &                              \\\hline
 %Hyades      &             Martin et al. (1998)      &               &  $ 650       \pm _{ 70}^{70       }$ &                               
\textbf{Praesepe} & \textbf{adopted in this work:}   &     185.5     &  $ 750\pm _{7}^{ 3 }$                &              0.16            \\
     &               \citet{bossini2019}     &               &                 $ 750\pm _{7}^{ 3 }$ &                              \\
     &            \citet{cantat_gaudin_2018} &     185.5     &                                      &                              \\
     &               \citet{gossage2018}     &               &                           590        &                              \\
     &             \citet{yen2018}           &     183       &  $ 794       \pm _{ 269}^{253     }$ &                              \\
     &\citet{netopil_metallicities_2016}     &               &                                      &               0.16           \\
     &             \citet{scholz2015}    &     187       &                           832        &                              \\
     &             \citet{boesgaard2013}   &               &                                      &               0.12           \\
     &             \citet{boudreault_astrometric_2012}  &     160       &                           630        &                              \\
     &            \citet{salaris_age_2004}     &     175       &                           650        &                              \\\hline
\textbf{Ruprecht 147} & \textbf{adopted in this work:}    &     305       & $ 2650      \pm _{ 380}^{380     }$  & $ 0.08          \pm 0.07  $  \\
     &             \citet{bragaglia2018}   &               &                                      &  $ 0.08          \pm 0.07  $ \\
      &           \citet{cantat_gaudin_2018} &     305       &                                      &                              \\
      &             \citet{gaia_dr2_2018_hrd} &     309       &  $ 1995      \pm _{ 257}^{404     }$ &                              \\
     &             \citet{torres2018}     &     283       &  $ 2650      \pm _{ 380}^{380     }$ &                              \\
    &             \citet{curtis2016}\tablefootmark{b} &               &                                      &  $ 0.10          \pm 0.02  $ \\
     &           \citet{scholz2015}    &     270     &                           1953       &                              \\
     &             \citet{curtis2013}    &     300       &  $ 3125      \pm _{ 125}^{125     }$ &  $ 0.07          \pm 0.03  $ 
\\\hline  
\textbf{M67} & \textbf{adopted in this work:}        &     908       & $3639  \pm _{ 17}^{17      }$        & $ -0.102         \pm .081  $ \\
          &             \citet{bossini2019}       &               &   $3639  \pm _{ 17}^{17      }$      &                              \\
          & \citet{netopil_metallicities_2016}    &               &                                      &               0.03           \\
          & \citet{barnes_rotation_2016}    &               &          $ 4200      \pm _{ 700}^{700}$                             &             \\

          &             \citet{scholz2015}    &               &  $ 3428      \pm _{ 72}^{147      }$ &                              \\
          &           \citet{conrad2014}     &               &                                      &  $ -0.102         \pm .081  $ \\
          &           \citet{dias_fitting_2012}      &     908       &                           4300       &                              \\
          &             \citet{onehag2011}    &     880       &                           4200       &               0.02           \\
          & \citet{salaris_age_2004}           &                   & $ 4300      \pm _{500}^{500}$    &       $0.02\pm 0.06$               \\\hline
\end{tabular}

\tablefoot{
\tablefoottext{a}{Bossini et al. (2019) noted some caveats for their determination of ages of young clusters, for which they used Gaia DR2 photometry for isochrone fitting.}
\tablefoottext{b}{Curtis (2016) reanalysed HIRES spectra using an improved spectroscopic method as compared to Curtis et al. (2013).
\tablefoottext{c}{We did not adopt a mean value for the Hyades distance because the cluster members are on average closer than 50 pc.}}
}
\end{table*}

%The identical ages for Paresepe in Yen+ and Kharchenko+ are coincidental
%--------------------------------------------------------------------
%-------------------------------------------------------------------
\section{Solar system object detection with SkyBot}
\label{app:skybot}
K2 photometry was heavily affected by crossings of Solar System Objects (SSO), which consequently produced mimicking flare profiles in the light curves~\citep{szabo2015}. Independently, two of us visually verified the SSO-nature of 42 flaring light curve events by inspecting their corresponding TargetPixelFiles (TPF) using \texttt{lightkurve}~\citep{lightkurve2018}. Except for one occasion, we found all events to be known SSOs, both in the visual survey and automatically using SkyBot~\citep{berthier2016}. Subsequently, we ran SkyBot on the remaining catalog of flares to remove other false positive events caused by known SSOs. Visual inspections of TPFs were carried out for any SSO detected within 30'' of its target star at mid-event time. This yielded additional 103 contaminated signals distributed among three subgroups, (i) 77 unambiguous SSO-contaminants verified through \texttt{lightkurve} (ii) 18 faint SkyBot-detections ($V\approx 20.4-23.8$) within $\sim3$ pixels from the respective target star but visually unresolvable in the TPF, (iii) 8 occurrences in which both flare and SSO-crossing occurred during the given flare duration stated in the catalog. The latter should be dealt with caution due to a high risk of inaccurate catalog information altered by the presence of SSOs. Despite of discarding a total of 144 flare events, we expect additional contaminants by unknown SSOs to hide in the current sample. This issue can partly be resolved by animating each TPF as demonstrated by Kristiansen et al. (in prep.). 
\section{Modified maximum likelihood estimatator}
\label{sec:app:MMLE}
As a means to arrive at results efficiently, and as consistency check to the method derived from~\citet{wheatland_flaresbayes_2004} we fitted $\alpha$ to the FFDs in $ED$ and $E_\mathrm{Kp}$ space using a modified maximum likelihood estimatator (MMLE,~ \citealt{maschberger2009}). The logarithm of the likelihood function $\mathcal{L}$ that had to be maximized was given by the authors in Eq. (8) in their manuscript:
\begin{equation}
\log \mathcal{L} = n \log (1-\hat{\alpha})-n \log\left(x_\mathrm{max}^{1-\hat{\alpha}}-x_\mathrm{min}^{1-\hat{\alpha}}\right) - \hat{\alpha} \displaystyle\sum_{i=1}^{n}\log x_i
\label{eqn:MLE}
\end{equation}
where $x_i$, $x_\mathrm{max}$, and $x_\mathrm{min}$ were the detected flare energies, and the upper and lower limits for detection, respectively. $n$ was the total number of flares. The estimate for $\alpha$ would be biased in practice because the value used for $x_\mathrm{max}$ would be the maximum energy that was measured, and not the underlying upper limit. The stabilization transformation suggested by the authors (Eq. (12) in \citealt{maschberger2009}) was then applied to the solution for $\alpha$ to account for this bias:
\begin{equation}
\alpha = 1 + \dfrac{n}{n-2}(\hat{\alpha} - 1)
\label{eqn:MLE_stabilize}
\end{equation}
Using the MMLE method on the full sample of flares in $E_\mathrm{Kp}$ and $ED$ space we obtained $\alpha_\mathrm{erg}=$\,1.97 and $\alpha_\mathrm{s}=$\,2.06\unskip, respectively, indicating a marginally flatter power law than the adopted \citet{wheatland_flaresbayes_2004} model ($\alpha_\mathrm{erg}=$\,1.98, $\alpha_\mathrm{s}=$\,2.1).
%--------------------------------------------------------------------
\end{appendix}
\end{document}